\documentclass[twocolumn,twocolappendix,]{aastex631} 
\usepackage{multirow}
\usepackage{gensymb}
\usepackage{float}
\usepackage{subfigure}
\usepackage{threeparttable}
\usepackage{amsmath}
\usepackage{color}
\usepackage{mathrsfs}
\usepackage{textcomp, gensymb}
\newcommand{\HI}{H\,\textsc{i}}
\newcommand{\OIII}{[O\,\textsc{iii}]}

\newcommand{\sersic}{{S\'ersic}}

\shorttitle{Low Star Formation Efficiency in Quiescent Galaxies}
\shortauthors{Li et al.}

\begin{document}
\title{The Origin of the Gas and Its Low Star Formation Efficiency in Quiescent Galaxies}

\correspondingauthor{Yang A. Li; Zhao-Yu Li}
\email{yangli\_anpin@outlook.com; lizy.astro@sjtu.edu.cn}

\author[0000-0002-3309-8433]{Yang A. Li}
\affil{Department of Astronomy, School of Physics and Astronomy, Shanghai Jiao Tong University, Shanghai 200240, P. R. China}
\affil{Shanghai Key Laboratory for Particle Physics and Cosmology, Shanghai 200240, P. R. China}
\affiliation{Kavli Institute for Astronomy and Astrophysics, Peking University, Beijing 100871, P. R. China}
\affil{Department of Astronomy, School of Physics, Peking University, Beijing 100871, P. R. China}

\author[0000-0001-6947-5846]{Luis C. Ho}
\affil{Kavli Institute for Astronomy and Astrophysics, Peking University, Beijing 100871, P. R. China}
\affil{Department of Astronomy, School of Physics, Peking University, Beijing 100871, P. R. China}

\author[0000-0002-4569-9009]{Jinyi Shangguan}
\affil{Kavli Institute for Astronomy and Astrophysics, Peking University, Beijing 100871, P. R. China}
\affil{Department of Astronomy, School of Physics, Peking University, Beijing 100871, P. R. China}

\author[0000-0001-5017-7021]{Zhao-Yu Li}
\affil{Department of Astronomy, School of Physics and Astronomy, Shanghai Jiao Tong University, Shanghai 200240, P. R. China}
\affil{Shanghai Key Laboratory for Particle Physics and Cosmology, Shanghai 200240, P. R. China}

\author{Yingjie Peng}
\affil{Kavli Institute for Astronomy and Astrophysics, Peking University, Beijing 100871, P. R. China}
\affil{Department of Astronomy, School of Physics, Peking University, Beijing 100871, P. R. China}

\begin{abstract}
Quiescent galaxies (QGs) typically have little cold gas to form stars. The discovery of gas-rich QGs challenges our conventional understanding of the evolutionary paths of galaxies. We take advantage of a new catalog of nearby, massive galaxies with robust, uniformly derived physical properties to better understand the origin of gas-rich QGs. We perform a comparative analysis of the cold interstellar medium and star formation properties of carefully matched samples of galaxies with varying degrees of star formation activity and gas richness. QGs with different gas content have virtually identical morphological types, light concentration, mass-size relation, stellar age, dark matter halo mass, and black hole activity. The only distinguishing characteristic is the environment. Gas-rich satellite QGs reside in a lower-density environment than their gas-poor counterparts, as a consequence of which they manage to retain their gas and experience a higher probability of cold gas accretion or gas-rich mergers. The environmental densities of central QGs are similar regardless of their gas content. We suggest that the cold gas resides mainly in the outskirts of the gas-rich QGs, where bars, if present, cannot transport it inward efficiently to fuel central star formation. The prominent bulges in gas-rich QGs stabilize the cold gas from fragmentation and leads to low star formation efficiency.
\end{abstract}


\section{Introduction}

The accretion and fragmentation of cold gas in galaxies fuel their star formation, contributing to their continuous growth in stellar mass ($M_*$) over cosmic time. Under normal circumstances, the amount of cold gas in a galaxy, especially in molecular form, is closely related to its star formation rate (SFR), a trend conventionally captured empirically through a relation attributed to the seminal works of \cite{Schmidt1959} and \cite{Kennicutt1998ApJ}. Passively evolving, quiescent galaxies [QGs, here defined as galaxies with $\mathrm{log}\,{\rm (sSFR\,/\,{\rm yr}^{-1})\equiv \mathrm{log}\,(SFR}/M_*)< -11.5$], those in which star formation has subsided, are usually depleted in cold gas ($M_{\rm gas}\lesssim 10^{9.3}\,M_{\odot}$; see Section~\ref{sec_match}). QGs can rekindle their star formation if their gas reservoir is replenished through external gas accretion, gas-rich mergers, or the infall of cooled halo gas \citep{Kaviraj2009, Rowlands2018}. However, in recent years, the discovery of a class of passive galaxies buckles conventional wisdom:  QGs endowed with as much cold gas as star-forming galaxies (SFGs) but that lie markedly below the Kennicutt-Schmidt relation on account of their low star formation efficiency ($\rm SFE\equiv SFR$/$M_{\rm gas}$). 

Examples of gas-rich QGs have been recognized for some time, particularly at higher stellar masses \citep{Pickering1997, Morganti2006, Young2009, Donovan2009, Gereb2018}. In recent years, attention has grown with the availability of large-area surveys that have led to an even wider appreciation of the existence of passive galaxies with unexpectedly abundant dust and cold (atomic and molecular) gas \citep{Parkash2019, Guo2020, Wang2022, Sharma2023, Lee2024,LiXiao2024}. Amassing Herschel observations over $\sim 160\,\rm deg^2$, \citet{Lesniewska2023} found a population of passive elliptical galaxies that nonetheless has significant far-infrared emission. They suggested that in these systems SFR declines faster with age than the dust mass. In fact, a substantial fraction of local quiescent early-type galaxies have detectable CO emission, and hence molecular gas \citep{Young2009, Young2011}. Interferometric observations reveal that the molecular gas is mostly confined to kpc-scale disks \citep{Alatalo2013, Boizelle2017}, which, on account of their location in the center of a massive spheroid, are stable against gravitational fragmentation and hence have low SFE (e.g., \citealt{Davis2014, Boizelle2017}). In their comprehensive study of high-mass ($M_*>10^{10.6}\,M_\odot$), central disk galaxies, \citet[][]{Zhang2019} find that the critical factor that controls star formation activity is the molecular gas, not atomic gas (but see \citealt{Cortese2020}). Relative to a matched control sample of SFGs, passive systems are markedly depleted in molecular gas but have a surprisingly similar reservoir of \HI, in agreement with the results from \citet{Hunt2020}. \cite{Zhang2021} confirmed that in the massive central disk galaxies, the \HI\ disks rotate as regularly as in the main-sequence disk galaxies, independent of the SFR indicator, likely due to the stable but low surface density \HI\ ring at large radii or the ejective mode active galactic nucleus (AGN) feedback.

Star formation in a galaxy can be quenched by various means. The most direct way is simply to deprive it of gas, for instance through ``cosmological starvation'' \citep{Feldmann2015} or galaxy harassment \citep{Moore1996}. Gas can also be removed by internal mechanisms such as AGN feedback \citep{Dubois2013, Weinberger2017} or externally through ram pressure stripping \citep{Gunn1972}. That is not the issue in hand. The galaxies under consideration have no shortage of gas. We want to know why, {\it in spite}\ of being endowed with gas, they are not amenable to forming stars. A central mass concentration, in the form of large stellar spheroid, would stabilize the gas disk against fragmentation and hence prevent star formation \citep{Martig2009}. Morphological quenching offers a plausible mechanism to account for gas-rich QGs \citep{Lesniewska2023}, although the lack of clear differences in the bulge-to-total ratios of the gas-rich and gas-poor QGs casts some doubt on the viability of this picture \citep{LiXiao2024}. Morphological quenching might also operate in tandem with feedback processes \citep{Lee2024}. 

The origin of the gas also presents a puzzle. \citet{Sharma2023} argue that external accretion supplies the gas to \HI-rich but low-SFR galaxies. This process can produce the counter-rotating \HI\ disks seen in \HI-excess galaxies \citep{Gereb2018} and also account for gas-enriched quiescent S0s that will ultimately be rejuvenated \citep{Rathore2022}. H~I-rich galaxies of low star formation activity are mostly central or isolated galaxies in low-mass halos, and they suffer little from strong environmental effects that would otherwise prevent them from retaining their gas \citep{LiXiao2024}.

This work investigates the possible mechanisms of gas enrichment and SFE suppression by performing a comprehensive statistical study of a large, well-defined sample of nearby galaxies with uniformly measured global SFRs and total cold gas masses, using carefully matched samples of galaxies with varying degrees of star formation activity and gas richness. Section~\ref{sec_sample} introduces the sample and parameters used in the paper. We compare the physical properties of the QGs with high and low gas content and present our results in Section~\ref{sec_result}. Section~\ref{sec_discussion} discusses possible physical processes responsible for reducing the SFE of gas-rich QGs. Our main conclusions are summarized in Section~\ref{sec_summary}. This paper assumes a cosmology with $H_0=70\,\rm km\,s^{-1}\,Mpc^{-1}$, $\Omega_m = 0.3$, and $\Omega_{\Lambda}=0.7$, and $M_*$ and SFR are based on the stellar initial mass function of \citet{Chabrier2003}.

\begin{figure*}
\centering
\includegraphics[width=0.95\textwidth]{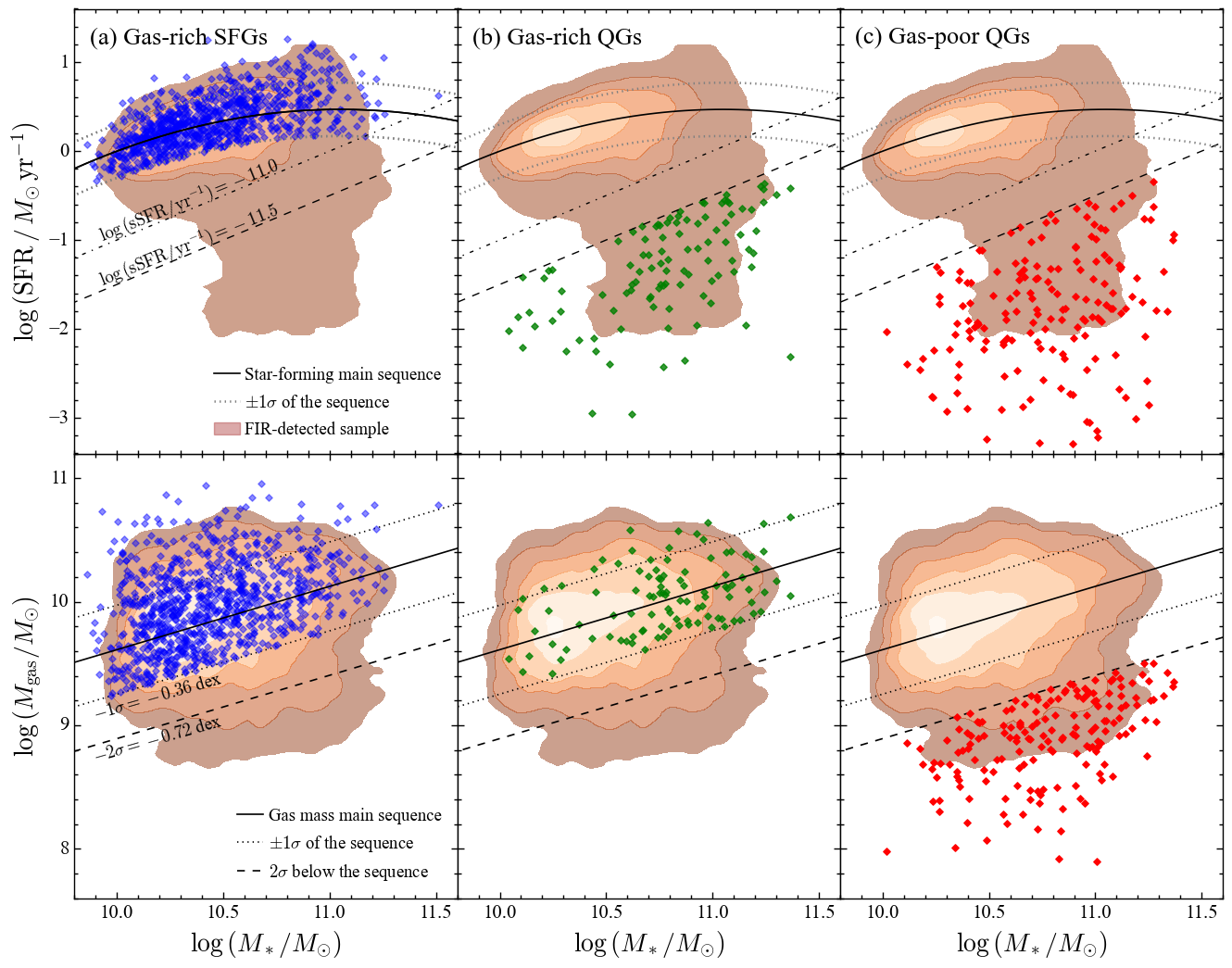}
\caption{Distributions of $M_*$ versus SFR (top) and $M_{\rm gas}$ (bottom) of (a) gas-rich SFGs, (b) gas-rich QGs, and (c) gas-poor QGs. The orange contours represent the FIR-detected sample defined in Section~\ref{sec_sample}. The black solid and dotted lines in the top panels show the best-fit star-forming main sequence and the $\pm 1\,\sigma$ range of the parent sample \citep{Li2023ApJS}. Dashed lines mark the levels of sSFR that separate star-forming, green valley, and QGs. The black solid and dotted lines in the bottom panels show the best-fit gas mass main sequence and $\pm 1\,\sigma$ range of the parent sample \citep{Li2023ApJ}. The dashed line indicates the $-2\,\sigma$ level from the gas mass main sequence, below which the gas content is much lower compared to that of SFGs.}
\label{fig:sample}
\end{figure*}

\section{Samples and Their Statistical Properties}\label{sec_sample}

The parent sample in this work consists of low-redshift ($z<0.11$), high-mass ($M_* > 10^{10}\,M_{\odot}$) galaxies contained in the Herschel/SPIRE survey \citep{VieroHerS2014} of the Sloan Digital Sky Survey (SDSS) Stripe~82 region \citep{Annis2014, Jiang2014}. The 17 bandpasses used in the spectral energy distribution (SED) analysis include the far-ultraviolet (FUV) and near-ultraviolet (NUV) bands from the Galaxy Evolution Explorer (GALEX; \citealt{Martin2005GALEX}), the optical $u$, $g$, $r$, $i$, and $z$ bands from SDSS, the near-infrared $J$, $H$, and $K_s$ bands from the Two Micron All Sky Survey (2MASS; \citealt{Skrutskie2006}), the mid-infrared W1, W2, W3, and W4 bands from the Wide-field Infrared Survey Explorer (WISE; \citealt{Wright_WISE2010}), and the far-infrared (FIR) 150, 250, and 350~$\mu$m bands from Herschel/SPIRE \citep{Griffin2010}.

\subsection{Stellar Masses, Total Gas Masses, and SFRs} 
The physical properties of the sample, such as SFR, $M_*$, and cold dust mass, have been carefully estimated in \cite{Li2023ApJS} from energy-balanced SED fitting of the panchromatic photometry using the \texttt{CIGALE} code \citep{Boquien2019}. The fit provides a probability-weighted mean and standard deviation of the physical parameters. For clarity, we use the measured values with appropriate errors (see \citealt{Boquien2019} for details), even though for some QGs the mean is less than 3 times the error, instead of adopting a $3\,\sigma$ upper limit. We emphasize that the two approaches are statistically consistent. As derived in \citet{Li2023ApJ, Li2023ApJS}, fitting the relationship between the logarithmic SFR and stellar mass with a polynomial function shows that SFGs follow a curved star-forming main sequence with a standard deviation ($1\,\sigma$) of 0.3~dex, while the gas content of gas-rich galaxies increases linearly with stellar mass along the gas mass main sequence with a standard deviation of 0.36~dex.
Our cold gas masses derive from the cold dust measurements, assuming that the gas-to-dust mass ratio ($\delta_{\rm GDR}$) is a function of gas-phase metallicity \citep[$Z_{\rm gas}$;][]{Magdis2012}. As an important constituent of the cold interstellar medium, dust is considered a reliable tracer of the total cold gas content in both star-forming \citep{Scoville2016} and QGs \citep{Magdis2021}, as well as in galaxies with significant AGN activity \citep{Shangguan2018, Shangguan2020}. We caution that the conversion between dust mass and total gas mass may be somewhat uncertain in QGs because the correlation between the gas-to-dust mass ratio and metallicity was derived based on main-sequence galaxies \citep{Magdis2012}, which may not be applicable to all QGs \citep{Lorenzon2025}. Judging from the stellar mass–metallicity relation of local galaxies \citep{Kewley2008}, our sample of QGs should track the $\delta_{\rm GDR}-Z_{\rm gas}$ relation of SFGs, and we expect that their gas masses should be reasonably well estimated from the dust \citep{Magdis2021}, with a typical uncertainty of $\sim 0.45$ dex \citep{Li2023ApJ}.

After excluding galaxies whose photometry is adversely affected by imperfectly removed nearby contaminating sources, the parent sample in \cite{Li2023ApJS} contains over 2500 low-redshift, massive galaxies. As shown in \cite{Li2023ApJS}, correctly measuring total dust mass requires detection in at least one FIR band. Therefore, we further select galaxies with signal-to-noise ratio $\rm S/N > 2$ in one or more Herschel/SPIRE bands, which yields a final FIR-detected sample of 1723 galaxies with reliable SFR and cold gas mass measurements.

\begin{figure}[t]
\centering
\includegraphics[width=0.45\textwidth]{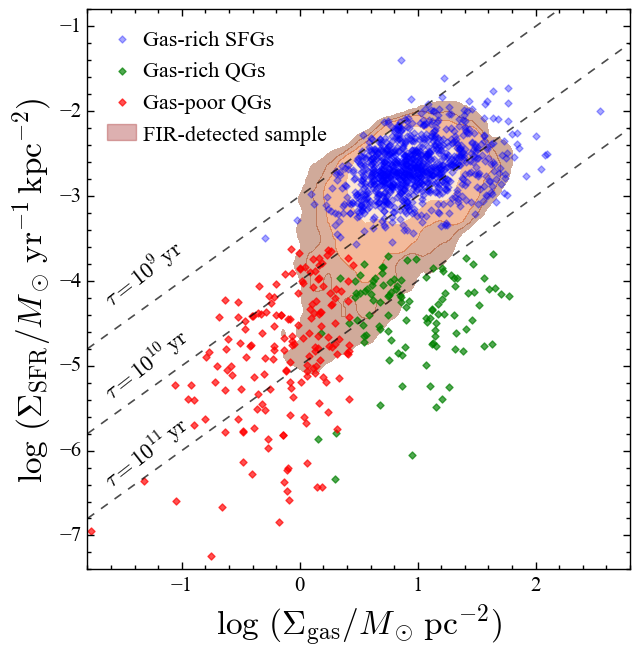}
\caption{The three samples of galaxies on the $\Sigma_{\rm SFR}-\Sigma_{\rm gas}$ diagram, overplotted in orange contours with the number density of galaxies in the parent sample of \cite{Li2023ApJS}. The dashed grey lines represent different gas depletion times ($\tau$). }
\label{fig:KSrelation}
\end{figure}

\begin{figure*}
\centering
\includegraphics[width=0.95\textwidth]{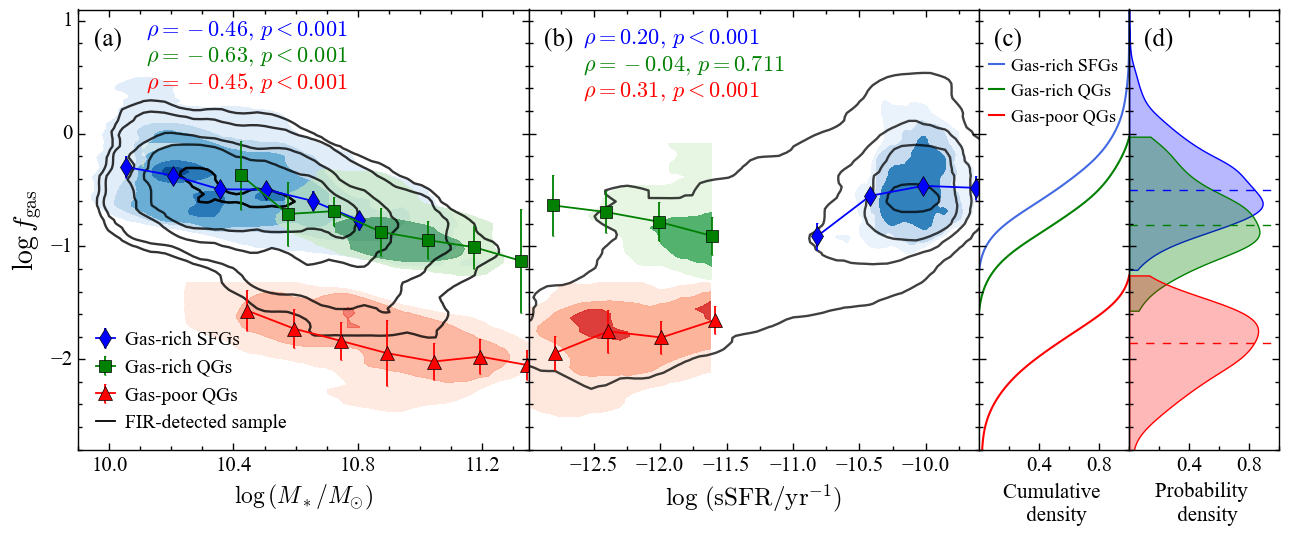}
\caption{Gas fraction ($f_{\rm gas}$) as a function of (a) $M_*$ and (b) sSFR for gas-rich SFGs (blue shaded contours), gas-rich QGs (green shaded contours), and gas-poor QGs (red shaded contours). Binned medians of those samples are shown as large symbols; the error bars of the symbols show the uncertainties of the median values, calculated by $3\,\sigma / N$, where $\sigma$ is the standard deviation of the distribution and $N$ is the number of galaxies in each bin of $M_*$ or sSFR. Black contours show the number density distribution of the parent sample in \cite{Li2023ApJS}. The normalized cumulative density and Gaussian-kernel probability density function for each sample are given in panels (c) and (d), respectively; the dashed lines in panel (d) mark the median values of $f_{\rm gas}$ for the three samples.}
\label{fig:Trend_fgas}
\end{figure*}

\subsection{Targets and Control Samples}\label{sec_match}
To investigate the nature of gas-rich galaxies with a low level of star formation activity, we define gas-rich QGs as those with specific SFR ($\mathrm{sSFR} = \mathrm{SFR}/M_*$) less than $10^{-11.5}\,\mathrm{yr^{-1}}$, below the conventional demarcation for the green valley \citep{Schawinski2014, Ray2024, Phillipps2019} but that otherwise are endowed with a total gas content typical of SFGs of similar stellar mass, as denoted by their adherence (within $\pm 1\,\sigma$) to the gas mass main sequence. The low sSFR cut ensures that our selected gas-rich QGs are fully quenched. We create two samples for comparison: (1) the gas-poor QGs, which apart from having $\rm sSFR < 10^{-11.5}\,\mathrm{yr^{-1}}$ are also required to lie at least $2\,\sigma$ (0.72~dex) below the gas mass main sequence; and (2) the gas-rich SFGs, which reside in the region $\rm sSFR > 10^{-11.0}\,\mathrm{yr^{-1}}$ and above $-1\,\sigma$ of the gas mass main sequence. These selection cuts are motivated by the larger uncertainties of the SFRs ($\sim 0.5$~dex) of the QGs and the larger uncertainties of the dust masses ($\sim 0.6$~dex) of the gas-poor galaxies \citep{Li2023ApJS}. The above criteria define a total of 107 gas-rich QGs, 173 gas-poor QGs, and 897 gas-rich SFGs. The median uncertainties of the SFRs of these three populations are 0.38, 0.68, and 0.12~dex, respectively, and for the dust masses, the corresponding uncertainties are 0.24, 0.64, and 0.27~dex. Figure~\ref{fig:sample} shows the distributions of the three samples on the SFR$-M_*$ and $M_{\rm gas}-M_*$ diagrams.

\begin{figure*}
\centering
\includegraphics[width=0.48\textwidth]{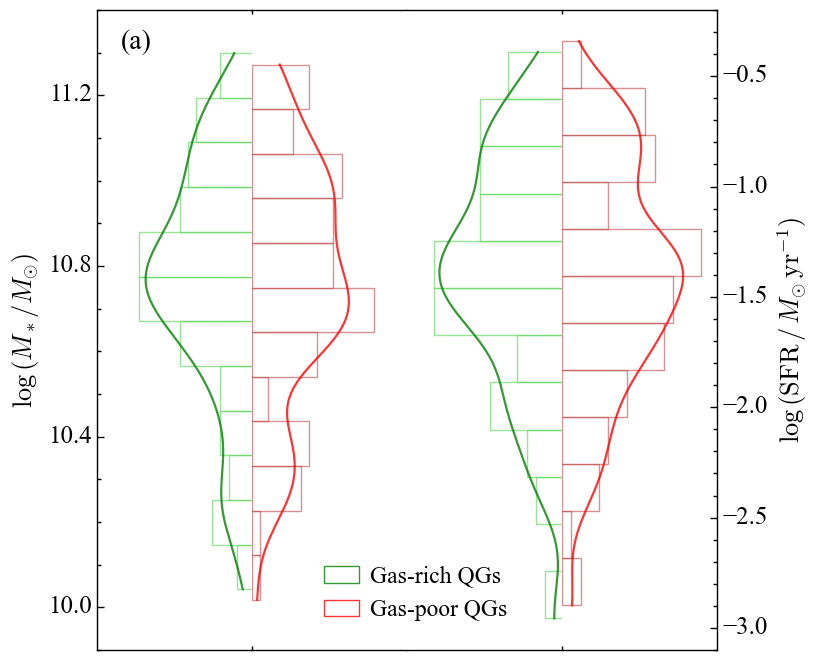}
\includegraphics[width=0.48\textwidth]{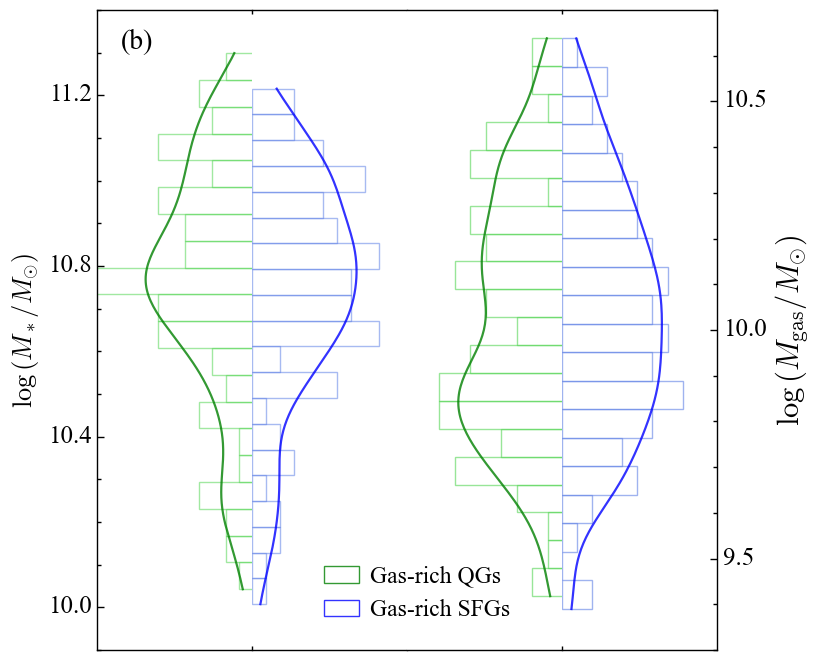}
\caption{Histograms of (a) $M_*$ and SFR, after matching gas-rich QGs (green) and gas-poor QGs (red), and (b) $M_*$ and $M_{\rm gas}$, after matching the gas-rich QGs (green) and SFGs (blue), overlaid by the kernel density estimate functions (solid curves). The histograms in each panel are consistent with each other.}
\label{fig:hist0}
\end{figure*}

\begin{figure*} 
\centering
\includegraphics[width=0.95\linewidth]{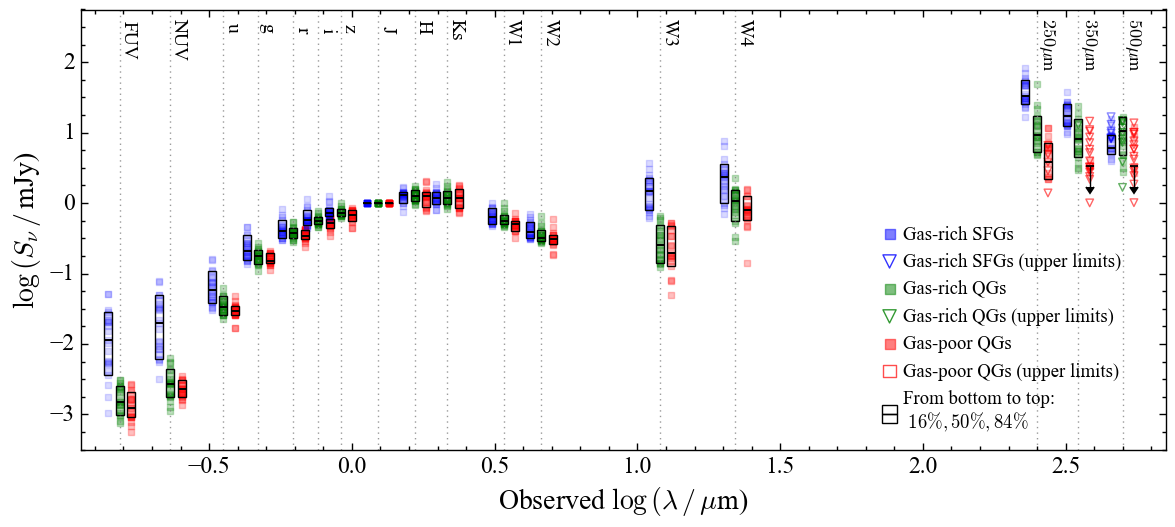}
\caption{Observed SEDs of gas-rich SFGs (blue), gas-rich QGs (green), and gas-poor QGs (red), normalized by the flux density in the 2MASS $J$ band ($1.2\,\mu$m). For each band, we also give the median and 16\% to 84\% range of the distribution of the sample. Vertical dotted lines denote the 17 bandpasses used in the SED fitting (Section~\ref{sec_sample}).}
\label{fig:SEDstack} 
\end{figure*}

To explore the SFE of the sample galaxies, we estimate the SFR surface density ($\Sigma_{\rm SFR}$) and the gas mass surface density ($\Sigma_{\rm gas}$), using the area $\pi a_{25}^2$, where $a_{25}$ is the semi-major axis of the galaxy's stellar disk, approximated by the isophotal radius at $25\,\rm mag\,arcsec^{-2}$ in the SDSS $r$-band \citep{Li2023ApJ}. As expected, normal, gas-rich SFGs track the standard Kennicutt-Schmidt relation with a zero point that corresponds to a gas depletion time scale of $\tau = 10^{9.62\pm0.37}\,\rm yr$ (Figure~\ref{fig:KSrelation}), typical of nearby galaxies on the star-forming main sequence (e.g., \citealt{Kennicutt1998araa, Bigiel2008}). By contrast, QGs collectively fall substantially {\it below}\ the mean locus of SFGs. Most notable are the gas-rich QGs, which despite having comparable values of $\Sigma_{\rm gas}$ nonetheless are characterized by a much longer gas depletion time scale of $\tau = 10^{11.27\pm0.58}\,\rm yr$. In other words, gas-rich QGs form stars with an efficiency nearly 2 orders of magnitude lower than that of SFGs. The gas-poor QGs form stars at similarly low levels ($\Sigma_{\rm SFR} \approx 10^{-5\pm1}\,M_\odot\,\rm yr^{-1}\,kpc^{-2}$), but they have less gas (by $\sim 1$\,dex) than their gas-rich counterparts. Thus, while their gas depletion times ($\tau = 10^{10.62\pm0.71}\,\rm yr$) are long, they are not quite as extreme as those of gas-rich QGs.

The relative statistical differences between the three galaxy samples also can be illustrated by plotting the gas mass fraction ($f_{\rm gas}=M_{\rm gas}/M_*$) as a function of $M_*$ and sSFR (Figure~\ref{fig:Trend_fgas}). In all three samples, $f_{\rm gas}$ decreases systematically toward the high-mass end, with similar slopes. Gas-rich QGs (green curve) are generally more massive than the gas-rich SFGs (blue curve), and, by design, exhibit suppressed sSFR. The gap in gas content between the groups with high and low $f_{\rm gas}$ is imposed by construction: we chose the subsamples based on their position relative to the gas mass main sequence (Figure~\ref{fig:sample}) to accentuate their possible physical differences. The gas fraction of the SFGs is slightly higher than that of the gas-rich QGs, although there is a continuous range in $f_{\rm gas}$ between the SFGs and gas-rich QGs (Figure~\ref*{fig:Trend_fgas}a).

Our goal is to explore the intrinsic differences, including morphology, halo mass, and AGN activity, in the gas-rich QGs compared to other galaxies. 
In order to reduce the bias among the three galaxy groups introduced by the systematic differences in SFR, stellar mass, and gas mass fraction, we further refine the control samples by matching $M_*$ and SFR for the gas-rich and gas-poor quiescent samples on the one hand, and by matching $M_*$ and $M_{\rm gas}$ for the star-forming and the gas-rich quiescent samples on the other hand (Figure~\ref{fig:hist0}). For each member of the gas-rich QG sample, we select a unique (non-redundant) gas-poor QG that closely matches the stellar mass ($\Delta M_*<0.1$~dex) and level of star formation ($\rm \Delta \,SFR < 0.15$~dex) of the gas-rich counterpart. If no suitable counterpart can be found, we relax the star formation criterion to $\rm \Delta\,SFR < 0.3$~dex, which is comparable to the 0.38~dex uncertainty of the SFRs in the QGs. Gas-rich QGs for which no match can be found among the gas-poor QGs are excluded. We then proceed to match the gas-rich SFGs to the gas-rich QGs that have well-matched gas-poor counterparts. We require a stellar mass difference of 0.1~dex and a gas mass difference of $<0.15$~dex, with an extra tolerance of 0.15~dex in the gas mass. The final three samples each contains 83 galaxies, whose respective physical properties are well matched (Figure~\ref{fig:hist0}) according to the Kolmogorov–Smirnov test \citep{Press1992KS}. The results of this work, derived based on the three parameter-matched samples, are primarily applicable to galaxies within the specific property ranges we selected. Nonetheless, we verified that our main conclusions are insensitive to whether the samples are parameter-matched. 

To illustrate visually the intrinsic differences among the three galaxy samples, Figure~\ref{fig:SEDstack} shows the normalized SEDs of the groups. For the purposes of this illustration, we further isolate a narrow redshift range ($0.05<z<0.08$) and impose a stellar mass cut of $M_* > 10^{10.6}\, M_{\odot}$ to minimize bias induced by the K-correction and any potential evolutionary effects.

\begin{figure*}
\centering
\includegraphics[width=0.275\textwidth]{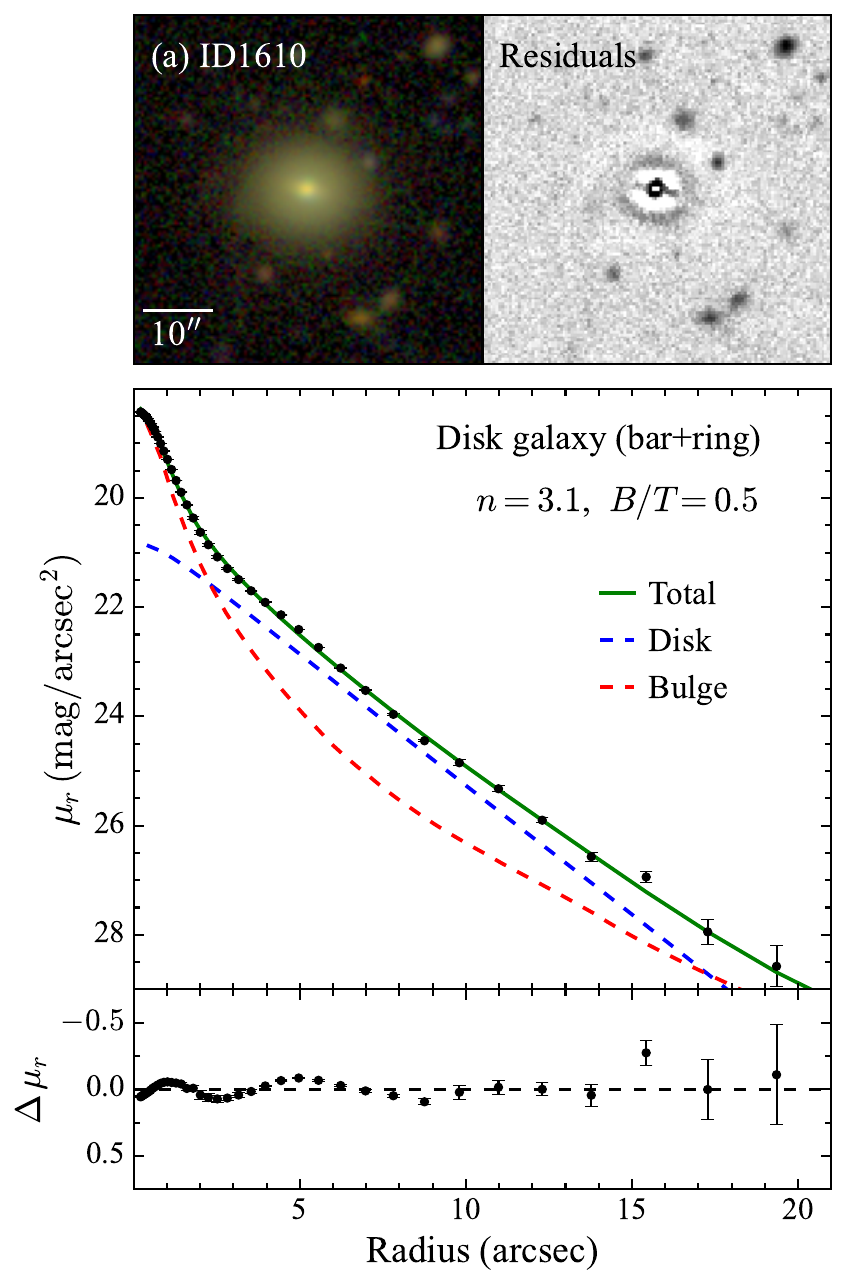} 
\includegraphics[width=0.2363\textwidth]{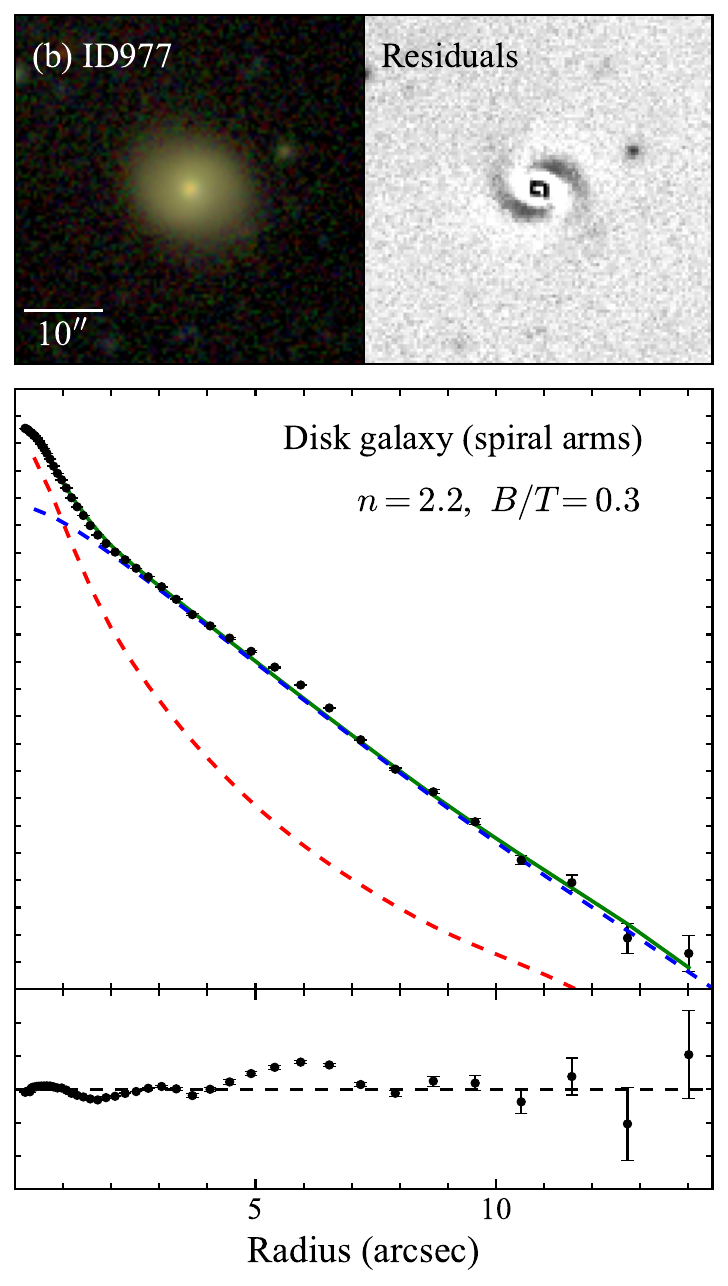}
\includegraphics[width=0.2363\textwidth]{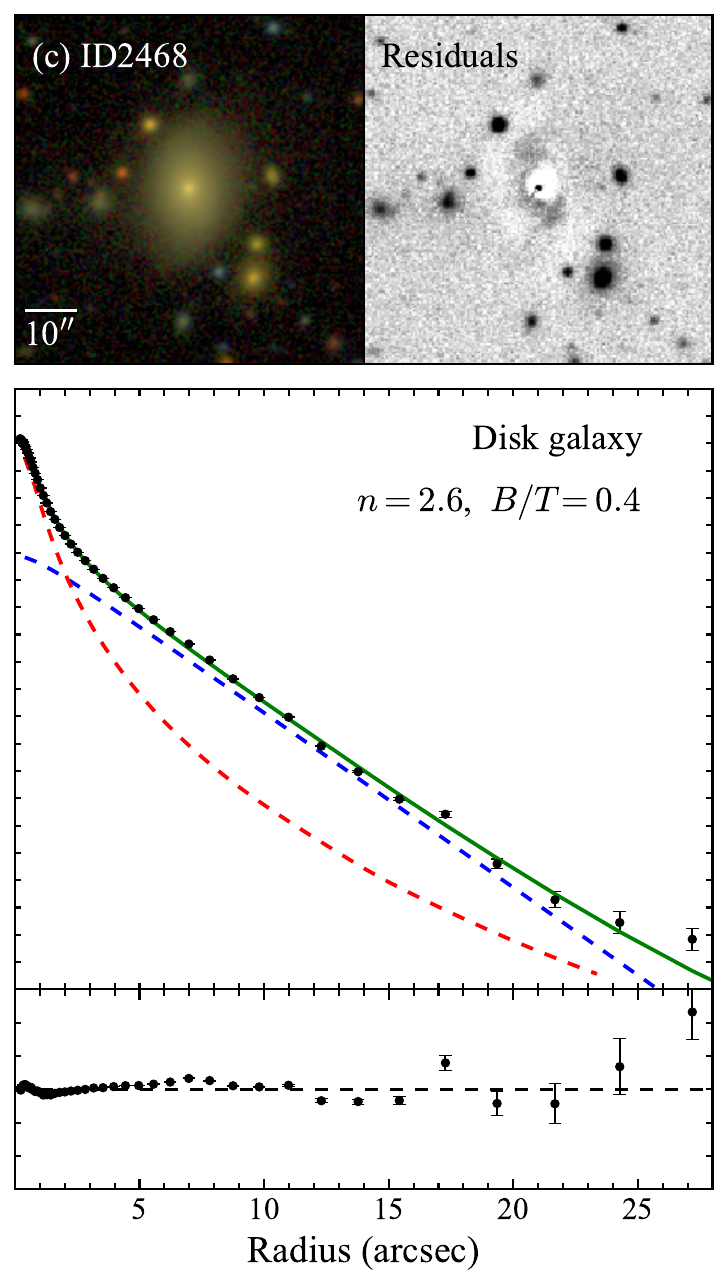}
\includegraphics[width=0.2363\textwidth]{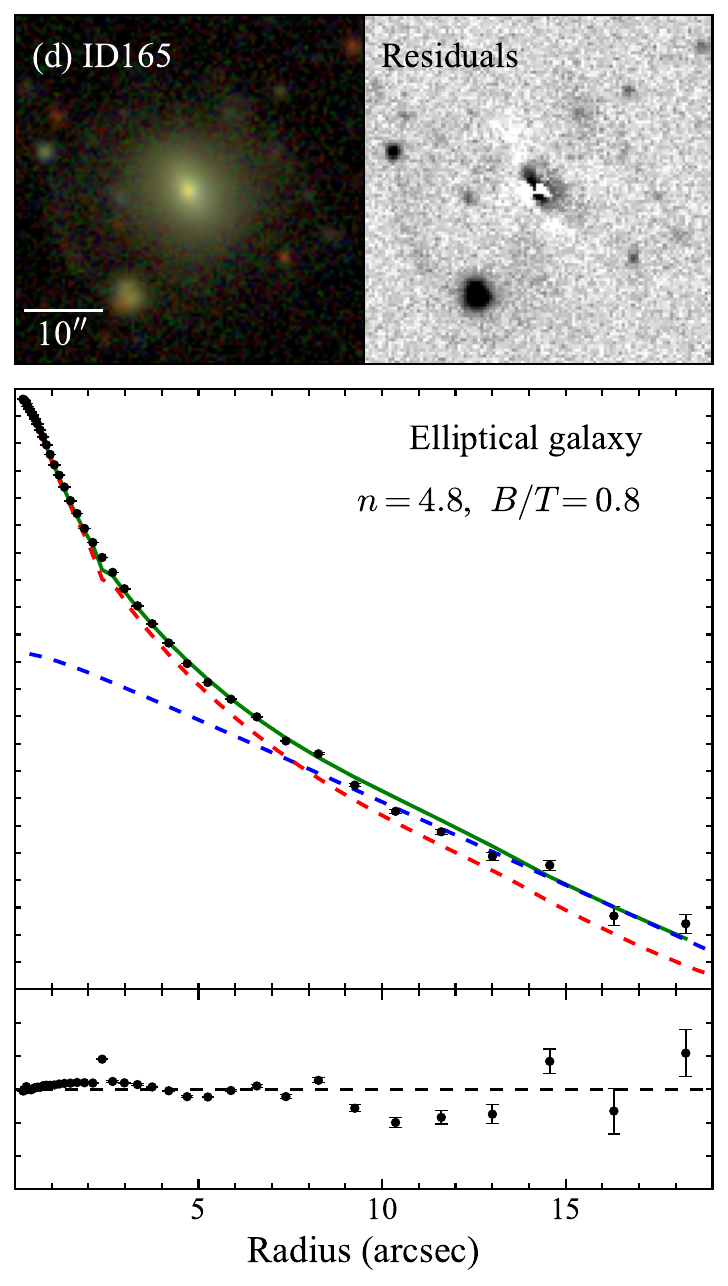}
\caption{ (Top) Examples of galaxy morphological assignment for four galaxies that cannot be easily classified by visual inspection from their original images, but whose nature is better revealed by the residual images: (a) a face-on disk galaxy with a bar-like feature; (b) a face-on disk galaxy with two spiral arms; (c) a face-on system with no substructure but that has $n<3.5$ and $B/T < 0.7$, which we regard as indicative of a disk galaxy; and (d) a galaxy deemed as an elliptical on the basis of $n\geq 3.5$ and $B/T\geq 0.7$.  (Middle) One-dimensional, background-subtracted  surface brightness profiles of the original images (black points with error bars) and of the bulge + disk best-fit models (red curve: bulge; blue curve: disk; green curve: total). (Bottom) Residuals of the surface brightness profiles between the original images and the best-fit models. }
\label{fig:image}
\end{figure*}

\begin{figure*}[h]
\centering
\includegraphics[width=0.95\textwidth]{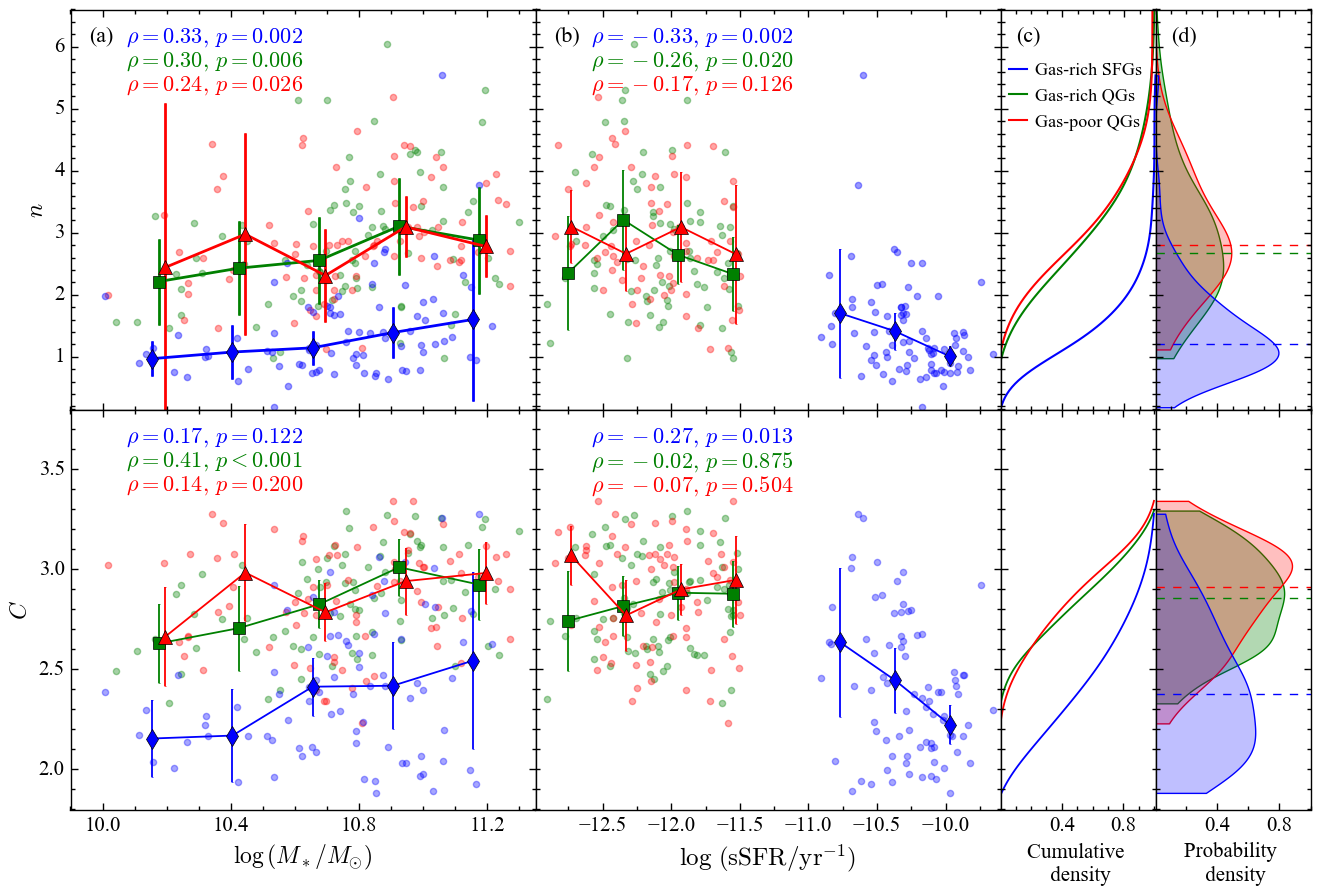}
\caption{S\'ersic index $n$ (top) and concentration index $C$ (bottom) as a function of (a) $M_*$ and (b) sSFR for gas-rich SFGs (blue), gas-rich QGs (green), and gas-poor QGs (red). The normalized cumulative density and Gaussian-kernel probability density function for each sample are given in panels (c) and (d), respectively; the dashed lines mark the corresponding median values of $n$ and $C$ for the three samples. The color-coded squares with error bars are the median values and their uncertainties, calculated by $3\,\sigma / N$, where $\sigma$ is the standard deviation of the distribution and $N$ is the number of galaxies in each bin of $M_*$ or sSFR. The Spearman rank correlation coefficient $\rho$ and $p$-value are given for each panel. }
\label{fig:n_total}
\end{figure*}


\begin{deluxetable*}{l ccccccc}
    \tabletypesize{\footnotesize}
    \tablecaption{ Statistical Properties of the Galaxy Samples \label{tab_statistics}} 
    \tablewidth{12pc}
    \tablehead{ 
    \colhead{  } & 
    \colhead{  } &
\multicolumn{2}{c}{ Gas-rich QGs }  &
\multicolumn{2}{c}{ Gas-poor QGs }  &
\multicolumn{2}{c}{ Gas-rich SFGs }  \\
    \colhead{ Property } & 
    \colhead{ Section } &
    \colhead{ Number } & 
    \colhead{ Percentage } & 
    \colhead{ Number } & 
    \colhead{ Percentage} & 
    \colhead{ Number } &
    \colhead{ Percentage} 
    \\ 
    \colhead{ (1) } & 
    \colhead{ (2) } & 
    \colhead{ (3) } & 
    \colhead{ (4) } & 
    \colhead{ (5) } & 
    \colhead{ (6) } &
    \colhead{ (7) } &
    \colhead{ (8) }
    }
    \startdata
    Disk & {\bf Sec.~\ref{sec_morphology} }           & 62 & $75\pm 13$ & 64 & $77\pm 13$ & 81 & $98\pm 15$\\
    Elliptical & {\bf Sec.~\ref{sec_morphology} }      & 21 & $25\pm  6$ & 19 & $23\pm  6$ &  2 & $ 2\pm  3$\\
    \hline
    Central & {\bf Sec.~\ref{sec_environment}  }        & 42 & $48\pm 10$ & 40 & $48\pm  9$ & 59 & $71\pm 12$\\
    Satellite & {\bf Sec.~\ref{sec_environment}  }        & 41 & $52\pm  9$ & 43 & $52\pm 10$ & 24 & $29\pm  7$\\
    \hline
    Star-forming & {\bf Sec.~\ref{sec_AGN}  }     &  1 & $ 1\pm  1$ &  1 & $ 1\pm  1$ & 40 & $48\pm  9$\\
    AGN & {\bf Sec.~\ref{sec_AGN} }             & 23 & $28\pm  7$ & 26 & $31\pm  7$ & 34 & $41\pm  9$\\
    Inactive & {\bf Sec.~\ref{sec_AGN} }        & 59 & $71\pm 12$ & 56 & $68\pm 12$ &  9 & $11\pm  4$\\
    \hline
    $f_{\rm radio}$  & {\bf Sec.~\ref{sec_AGN} } &  3 & $21\pm 14$ &  1 & $ 6\pm  6$ &  1 & $ 9\pm  9$\\
    \enddata
    \end{deluxetable*}

\section{Results}\label{sec_result}

Various processes may account for the characteristics of gas-rich QGs, including morphological quenching \citep{Martig2009, Michalowski2024}, AGN feedback \citep{Sharma2023, Dou2024}, and environmental quenching \citep{LiXiao2024}. We compare the three samples defined in Section~\ref{sec_sample} to identify possible quenching mechanisms and gain insight into the underlying causes of their observed differences. It is important to emphasize that after the sample-matching process, the three samples are no longer representative of their original populations. Consequently, the conclusions drawn in this work cannot be extrapolated directly to galaxies outside the specific dynamical ranges examined. However, when examining the same parameters in galaxy samples before matching their stellar mass, SFR, and gas mass, we observe results consistent with those presented in this section.

\subsection{Morphological Quenching}\label{sec_morphology}

Morphological quenching is the most commonly invoked explanation for curtailing star formation in spite of the availability of cold gas \citep{Martig2009}. Cold gas must be gravitationally unstable to form stars. \cite{Toomre1964} characterized the stability of a galactic gas disk in terms of a parameter that is proportional to the gas velocity dispersion and the depth of the gravitational potential. As galaxies with a deeper gravitational potential have a larger stellar velocity dispersion, higher gas velocity dispersion \citep{Ho2009b}, and increased gas shear \citep{Davis2022, Lu2024}, the spheroidal component of bulge-dominated disk (spiral and S0) galaxies and elliptical galaxies provides a natural barrier to cold gas fragmentation and collapse, and consequently leads to less efficient star formation than in disk-dominated systems \citep{Kennicutt1989}. Motivated by these considerations, we compare the relative incidence of disk versus elliptical morphological classifications between the gas-rich and gas-poor QGs, study the incidence of morphological subcomponents such as bars and spiral arms, and characterize their global structural parameters. Gas-rich SFGs are used as a reference for comparison.

We classify the overall galaxy morphology into two broad categories---disk and elliptical galaxies---based on the structural parameters derived from image decomposition, as well as visual inspection of substructures on model-subtracted residual images. The morphological classifications available from the Galaxy Zoo~2 project \citep{Willett2013} are not used in this work because of the well-known difficulty of visually distinguishing face-on S0s from ellipticals (e.g., \citealt{vandenBergh1989,Zhang2021}). We take advantage of the increased depth of the coadded images of the SDSS Stripe~82 region, which reaches a surface brightness limit of $\sim 28.5\,\rm mag\,arcsec^{-2}$ in the $r$ band \citep{Jiang2014, Fliri2016}.  The measurement of structural parameters such as the bulge-to-total flux ratio ($B/T$) or \cite{Sersic1968} index ($n$) is significantly improved using the coadded images \citep{Bottrell2019} compared to the single-epoch images \citep{Meert2015}. 

To secure a homogeneous set of structural parameters for the sample galaxies, we use \texttt{GALFIT} \citep{Peng2002GALFIT, Peng2010GALFIT} to perform a single-component, global fit as well as a double-component, bulge-disk decomposition of the coadded $r$-band images stacked by \citet{Fliri2016}. The catalog of \citet{Bottrell2019}, while a valuable resource of bulge-disk decomposition for the Stripe~82 region, only covers $\sim 80\%$ of the galaxies in our sample, and it does not give global fits. Two-component decomposition furnishes a quantitative metric ($B/T$) to gauge the relative flux contribution of the bulge, and in addition gives as a by-product: a model-subtracted residual image that can be used to identify morphological substructures such as spiral arms, bars, and rings, which are otherwise difficult to discern from the original images (e.g., \citealt{Ho2011,Deng2023}). For each galaxy, we create an image cutout with a minimum size of $200\arcsec \times 200\arcsec$ up to 4 times $a_{25}$. In order to detect and mask sources of different brightness and S/N in the various SDSS bands, we use the \texttt{Python} package \texttt{photutils} \citep{Bradley2020} to perform source detection on the stacked $gri$ images of \citet{Fliri2016}. We generate a mask image with a detection threshold of 1.5 times the sky standard deviation for at least five connecting pixels. The detected sources, except for the target, are masked using elliptical segments with geometric information (e.g., semimajor and semiminor axis, and position angle) derived using the function \texttt{detect\_sources} and a size magnification factor of 3. To separate target galaxies with close neighbors, we additionally use the function \texttt{deblend\_sources} to break up the original detection segments into smaller ones based on the saddle point between the flux peaks. \texttt{GALFIT} automatically produces a sigma image based on the gain and the number of images used for stacking\footnote{The weight images that accompany the coadded images generated by \citet{Fliri2016} provide the number of input frames contributing to the final stack.}. We fit the two-dimensional sky background using a second-order polynomial. In the GALFIT modeling, we adopt the composite, extended point-spread functions constructed from the Stripe 82 data set \citep{Infante2020}.

For the two-component decomposition, we fit the light distribution using a \sersic\ function with a fixed index of $n=4$ for the bulge plus an $n=1$ (exponential function) for the disk, limiting the solution such that the axis ratio $b/a>0.5$ for the former and $b/a>0.1$ for the latter. The bulges of nearby ($z \approx 0$) spiral and S0 galaxies span a broad distribution of S\'ersic indices between $n \approx 1$ to 6, peaking near $n = 2$ \citep{Gao2020}. Extensive experimentation reveals that while allowing $n$ to vary freely between 1 and 6 does not significantly influence $B/T$ [typical difference of $\Delta\,(B/T)<8\%$ compared to the case of fixing $n$ to 4], the fits are more susceptible to the influence of bars or nuclei, and leaving $n$ unconstrained produces residual images that are more ambiguous to interpret and less suitable for the identification of substructure.  

We determine the morphological types as follows. Galaxies with axis ratio $b/a < 0.5$, which comprise 27\% of the sample, are too flat to be consistent with most ellipticals \citep{Sandage1970, Ryden2001} and are classified automatically as (highly inclined or edge-on) disk systems. We regard the remaining less-inclined sources with $b/a \geq 0.5$ as disk galaxies if we notice substructures such as a bar, ring, or spiral arms on the model-subtracted residual images; 44\% of the sample fall in this category. Relatively face-on sources without obvious substructures (12\% of the sample) pose the greatest challenge because unbarred, featureless S0s can be easily confused with ellipticals \citep{YuSY2020}. Informed by the statistical properties of nearby galaxies \citep{Huang2013, Gao2018, Gao2019, Gao2020}, for present purposes, we classify such galaxies as ellipticals if they have $n>3.5$ or $B/T\geq 0.7$ and as disk galaxies if $n\leq 3.5$ and $B/T < 0.7$. Examples of morphological classification are given in Figure~\ref{fig:image}.  It is worth noting that non-axisymmetric low surface brightness features in galaxy outskirts, such as tidal features or streams, may affect the morphological fitting, especially in images that can reach a depth of $\gtrsim 30\, \mathrm{mag\,arcsec}^{-2}$ \citep{Abraham2014, Skryabina2024}. However, the Stripe 82 images reach a $3\sigma$ depth limit of $\sim 28.5\, \mathrm{mag\,arcsec}^{-2}$ (10$\times$10 arcsec) in the $r$-band, where the tidal features mainly dominate regions with radii between $a_{\rm 25}$ and $2\,a_{\rm 25}$ \citep{Li2023ApJS}. In this work, we perform the 2D morphological fitting up to radii exceeding $ 4\,R_{\rm 90}$, which are typically larger than $2\,a_{\rm 25}$. In such a large fitting region, the low surface brightness features are azimuthally averaged in the 2D fitting process and do not significantly affect the final 2D models.

\begin{figure*}[h]
\centering
\includegraphics[width=0.95\textwidth]{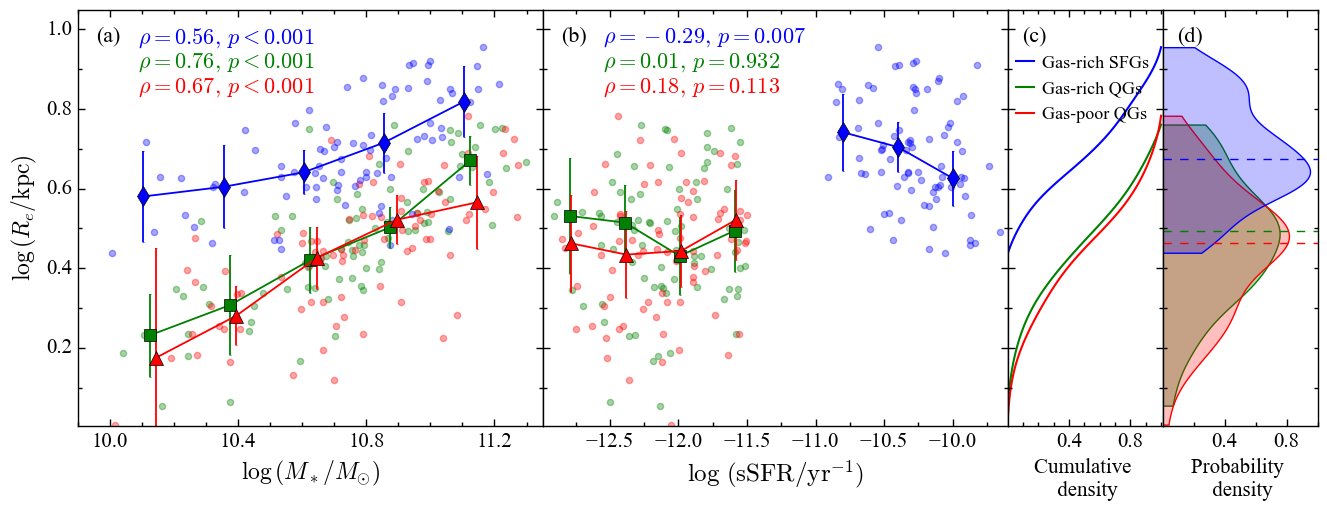}
\caption{Effective radius $R_e$ as a function of (a) $M_*$ and (b) sSFR for gas-rich SFGs (blue), gas-rich QGs (green), and gas-poor QGs (red). The normalized cumulative density and Gaussian-kernel probability density function for each sample are given in panels (c) and (d), respectively. Other conventions are similar to Figure~\ref{fig:n_total}.}
\label{fig:Re_total}
\end{figure*}

\begin{figure*}[t]
\centering
\includegraphics[width=0.95\textwidth]{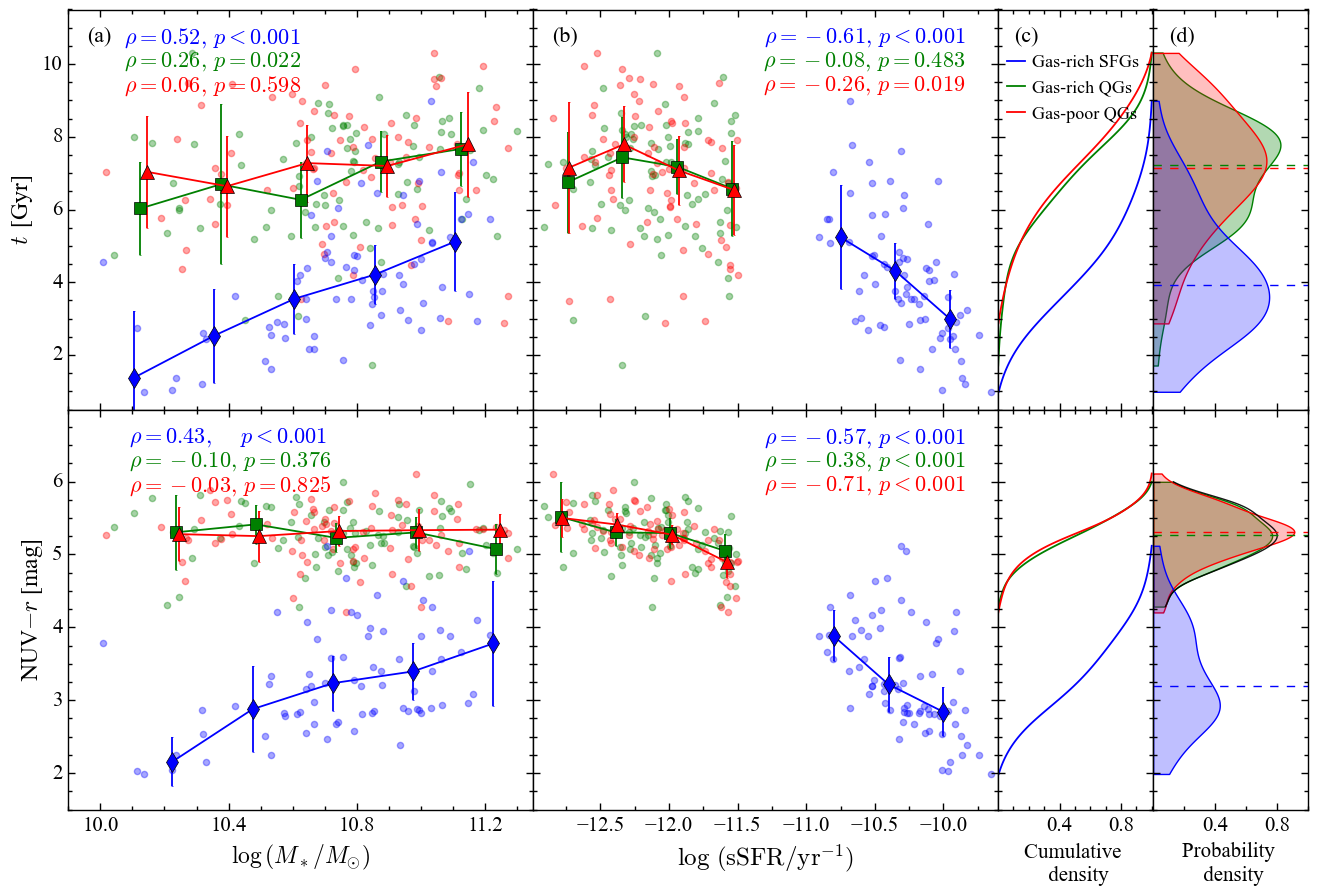}
\caption{Central stellar age $t$ (top) and global NUV$-r$ color (bottom) as a function of (a) $M_*$ and (b) sSFR for gas-rich SFGs (blue), gas-rich QGs (green), and gas-poor QGs (red). The normalized cumulative density and Gaussian-kernel probability density function for each sample are given in panels (c) and (d), respectively. Other conventions are similar to Figure~\ref{fig:n_total}.}
\label{fig:t_Ms}
\end{figure*}

Table~\ref{tab_statistics} lists the number fractions of disk and elliptical morphologies in the three samples. After matching the basic physical properties ($M_*$, SFR, $M_{\rm gas}$) in the three samples, we find that gas-rich and gas-poor QGs have similar constituents of disk ($75\%\pm 13\%$ and $77\%\pm 13\%$, respectively) and elliptical ($25\%\pm 6\%$ and $23\%\pm 6\%$, respectively) members. The high incidence of disks among QGs bears a strong similarity to the gas-rich SFGs, which are almost universally ($98\%\pm 15\%$) disky. This fraction is higher than the incidence of disks in previous samples of QGs studied through visual classification \citep{Bundy2010, Moffett2016}. Detailed, quantitative structural decomposition clearly is better able to distinguish between face-on S0 and elliptical galaxies. 

Figure~\ref{fig:n_total} presents two views of the variation of galaxy structure with stellar mass and sSFR. As anticipated, the QGs (green and red symbols) exhibit notably higher $n$ than the SFGs (blue symbols) at fixed $M_*$. The QGs have a comparable distribution of $n$, with median values of $2.7\pm 1.2$ for the gas-rich QGs and $2.8\pm 1.3$ for the gas-poor QGs, as illustrated in the cumulative and probability density functions (panels c and d), in contrast to a median value of $1.2\pm 0.8$ for the SFGs. All three samples show a mildly significant trend of $n$ increasing with $M_*$ (Spearman's rank order correlation coefficient $\rho>0.2$ and $p<0.05$), while the opposite trend, apart from the gas-poor QGs, holds for sSFR. Turning next to the concentration index $C=R_{90}/R_{50}$ as a measure of galaxy morphology, where $R_{90}$ and $R_{50}$ denote, respectively, the radius encompassing 90\% and 50\% of the SDSS Petrosian flux in the coadded $r$-band images \citep{Shimasaku2001}, the dependence of $C$ on $M_*$ and sSFR is weaker than the case for $n$, although the close similarity between the two types of QGs (median $C = 2.9\pm 0.3$ for the gas-poor and $2.9\pm 0.2$ for the gas-rich variety) and their difference with SFGs ($C=2.4\pm 0.3$) is obvious.  For reference, in the SDSS DR16 database, early-type galaxies at $z<0.08$ have $C>2.65$, while galaxies with $C\leq 2.65$ are considered late-type \citep{Kinyumu2024}. These results suggest that among quiescent galaxies neither their variation in gas content nor SFE can be ascribed to gross differences in galaxy morphology.

 The QGs in our study, irrespective of their gas content, delineate a stellar mass-size relation that is both steeper and offset below that of SFGs (Figure~\ref{fig:Re_total}).  Evidently, both groups of QGs experienced a similar assembly history, distinct from that of SFGs. Distinct from gas-rich SFGs, whose median effective radius shrinks with rising sSFR, the size of the QGs remains constant with sSFR at low redshift because their stellar mass growth is dominated by dry mergers instead of star formation \citep{vanDokkum2015}. QGs, be they rich or poor in gas, show similarly evolved stellar populations as indicated by their light-weighted age ($t$) estimated by \citet{Gallazzi2005, Gallazzi2006} from a likelihood analysis of the spectra in the central $3\arcsec$-diameter SDSS fiber. The same holds for the global NUV$-r$ colors, which are uniformly redder than those of SFGs (Figure~\ref{fig:t_Ms}).

\begin{figure}[t]
\centering
\includegraphics[width=0.47\textwidth]{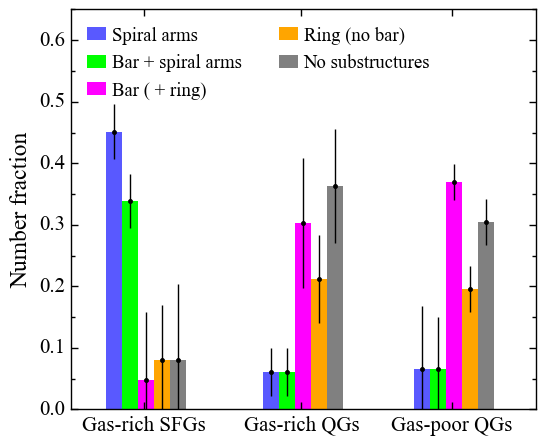}
\caption{Number fraction of disk galaxies with different substructures. Note that edge-on galaxies have been excluded. The gas-rich and gas-poor QGs have similar morphologies, dominated by S0 galaxies.}
\label{fig:barfrac}
\end{figure}

\subsection{Quenching by Spiral Arms or Bars} \label{sec_bar}

Do disk substructures, such as spiral arms and bars, play a role in suppressing star formation? Figure~\ref{fig:barfrac} shows the percentages of galaxies with different substructures, with the uncertainties given by Poisson statistics\footnote{The uncertainty of the percentage $f=N/M$ is denoted by $\delta f = f \sqrt{(\frac{\sqrt{N}}{N})^2 + (\frac{\sqrt{M}}{M})^2}$, where $N$ is the number of galaxies exhibiting a certain morphological classification and $M$ is the total number of galaxies in the sample of interest.}, for all low-inclination ($b/a\leq0.5$) disk galaxies whose model-subtracted residual images were examined in Section~\ref{sec_morphology}. Combining both unbarred and barred spirals into a single category reveals that only $\sim 12\%$ of the gas-rich QGs have spiral structure, nearly exactly the same as in gas-poor QGs ($13\%$), both significantly lower than the 79\% seen among SFGs. Thus, most QGs can be regarded as S0s, with a significant fraction hosting a bar or a ring. This is in agreement with the increasing bar frequency in massive disk galaxies in the quenching process \citep{Zhang2021}. The dearth of spiral structure in QGs is consistent with the notion that spiral arms help regulate galaxy-scale star formation \citep{Roberts1969, Kendall2015, YuSY2020, Yu2022}. 

\begin{figure*}[t]
\centering
\includegraphics[width=0.9\textwidth]{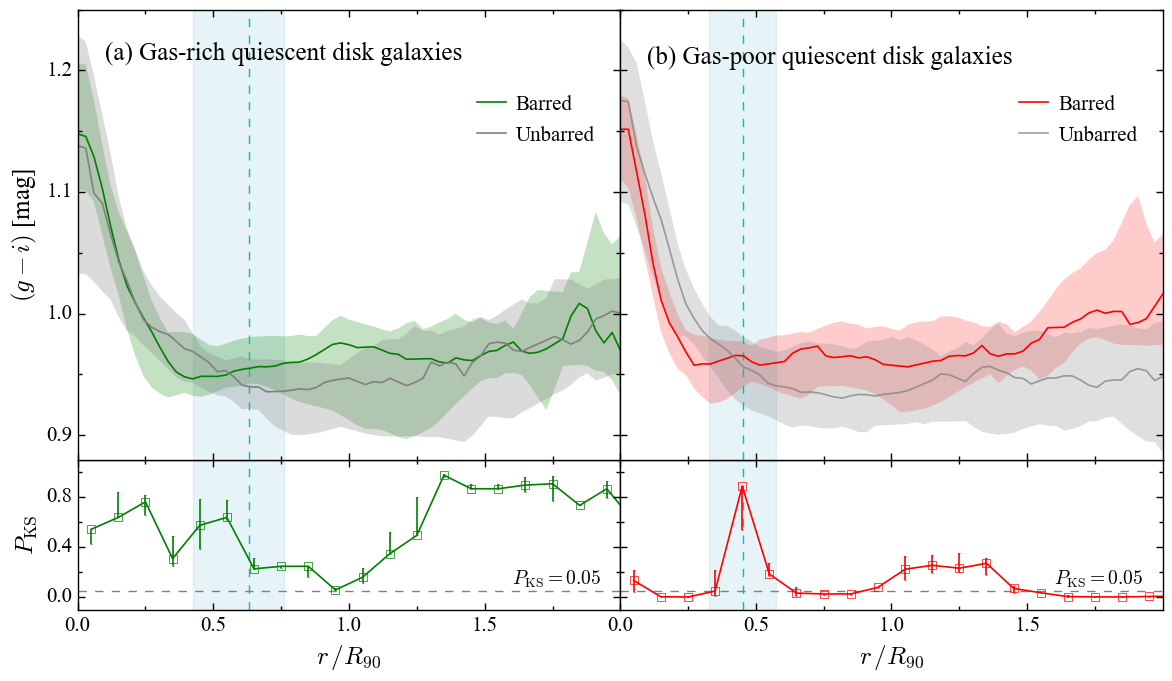}
\caption{Comparisons of the stacked $g-i$ color profiles of the barred and unbarred (grey) galaxies in the (a) gas-rich (green) and (b) gas-poor (red) quiescent disk galaxy samples. The solid curves on the top plots show the median of the stacked color gradients, with the shaded regions denoting the 16th and 84th percentiles.  The cyan vertical dashed lines with shaded regions denote the median, 16th, and 84th percentiles of $a_{\mathrm{bar}}$ of the barred galaxies. The bottom plots give the Kolmogorov-Smirnov probability $P_{\rm KS}$ for the comparison of the color profiles between the barred and unbarred samples. The grey horizontal dashed line indicates $P_{\rm KS}=0.05$, below which the two samples would be considered to be drawn from a statistically different parent distribution.}
\label{fig:bar_unbar}
\end{figure*}

From a theoretical point of view, a stellar bar exerts torque on the gas, induces gas inflow toward the central region of the galaxy (e.g., \citealt{Sormani2015, Fragkoudi2016}), and promotes the growth of the bulge and, if present, the supermassive black hole (BH) \citep{Kormendy2004}. However, we see no notable difference in bar fraction between gas-rich and gas-poor QGs, suggesting that bars have little effect on the gas content of these systems, at least as currently observed. Might there be traces of bar-induced secular evolution operating in the recent past?  We test this by comparing the age of the stellar population, as indicated by the $g-r$ color, of the barred and unbarred galaxies at different radii extending to $2\,R_{90}$. Figure~\ref*{fig:bar_unbar} presents the comparisons for both the gas-rich and gas-poor QGs.  We mark the radius ($a_{\mathrm{bar}}$) where the bar ellipticity reaches a maximum to guide the eye \citep{Marinova2007, Zhao2020}. Careful scrutiny of the $g-i$ color gradients of the two QG samples reveals that barred gas-poor QGs (Figure~\ref*{fig:bar_unbar}b) are generally redder at large radii ($r>R_{90}$) and bluer in the central regions ($r< 0.5\,R_{90}$) than their unbarred counterparts. This effect is absent from the gas-rich QGs (Figure~\ref*{fig:bar_unbar}a), as indicated by the Kolmogorov–Smirnov test probabilities ($P_{\rm KS}>0.05$) averaged in radial bins\footnote{At each radius, we perform a Kolmogorov–Smirnov test to compare the optical colors of the barred and unbarred samples, and then calculate the mean and standard deviation of $P_{\rm KS}$ of each radial bin.}. The bluer central colors of the barred gas-poor QGs suggest recent gas transport from the outer disk to the inner region, which triggered star formation in their center. The absence of radial color variations with the presence of a bar among gas-rich QGs may reflect the long timescale of secular evolution at low redshifts.

\begin{figure*}[t]
\centering    
\includegraphics[width=0.95\textwidth]{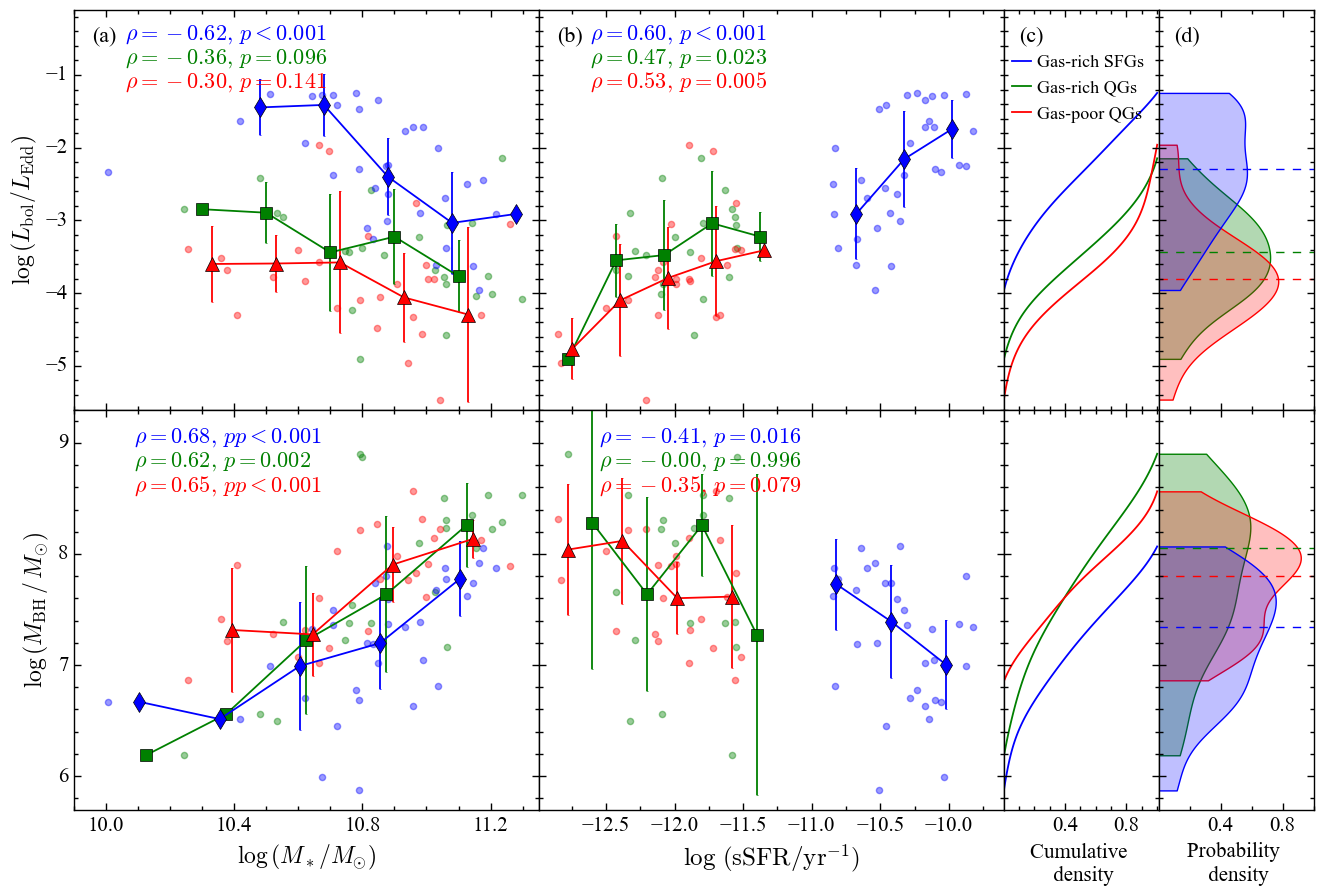}
\caption{Eddington ratio $L_{\rm bol}/L_{\rm Edd}$ (top) and BH mass $M_{\rm BH}$ (bottom) as a function of (a) $M_*$ and (b) sSFR for gas-rich SFGs (blue), gas-rich QGs (green), and gas-poor QGs (red). The normalized cumulative density and Gaussian-kernel probability density function for each sample are given in panels (c) and (d), respectively. Other conventions are similar to Figure~\ref{fig:n_total}.}
\label{fig:AGN_Ms}
\end{figure*}

\subsection{AGN Feedback Quenching}\label{sec_AGN}

AGN feedback can prevent star formation by injecting energy into the cold gas or expelling it from the galaxy \citep{Zinger2020}. We classify AGNs according to their optical diagnostic emission-line intensity ratios \citep{Kewley2006}, lumping together both narrow-line (type~2) AGNs and composite galaxies that contain both accretion-powered and star-forming activity, as well as broad-line (type~1) sources that are separately flagged as ``QSO'' in the SDSS catalog.  After matching the physical properties to create the target and control samples (Section~\ref{sec_match}), only one QSO remains, which is in the SFG sample. All the other AGNs in our samples of interest are type 2 or composite galaxies. Among the quiescent galaxies, $28\%\pm 7\%$ of the gas-rich group are spectroscopically classified as AGNs, essentially indistinguishable from the $31\%\pm 7\%$ found in the gas-poor members. Both are marginally lower than the SFGs, which have an AGN fraction\footnote{While it may seem paradoxical to classify a SFG as an AGN, note that we define SFGs in terms of their global sSFR (Section~\ref{sec_sample}), not emission-line properties. The host galaxies of spectroscopically classified AGNs have a broad range of star formation activity (e.g., \citealt{Ho2003, Zhuang2020}), and therefore it is not surprising that some SFGs are classified as AGNs according to their nuclear spectra.} of $41\%\pm9\%$ (Table~\ref{tab_statistics}). 

Following \citet{Kong2018}, we derive the bolometric luminosity ($L_{\rm bol}$) of the AGN from the extinction-corrected \OIII\ $\lambda 5007$ luminosity ($L_{\rm \OIII}$), assuming a constant scale factor of $L_{\rm bol}=600 \,L_{\rm \OIII}$ and internal extinction estimated from the Balmer decrement with an intrinsic value of H$\alpha$/H$\beta=3.1$ \citep{Halpern1983}. Galaxies that lack a reliable measurement of the Balmer decrement were assigned H$\alpha$/H$\beta=4.0$, the median value observed in nearby, luminous type~2 AGNs \citep{Kong2018}. The Eddington luminosity $L_{\rm Edd}=1.3\times 10^{38}\,(M_{\rm BH}/M_{\odot})$, with the BH mass ($M_{\rm BH}$) estimated from the local $M_{\rm BH}-\sigma_*$ relation using central stellar velocity dispersions ($\sigma_*$) from the MPA-JHU catalog. \citet{KormendyHo2013} only provide the $M_{\rm BH}-\sigma_*$ relation for classical bulges and elliptical galaxies. To facilitate analysis of galaxies with a wider range of morphological types, \citet{She2017} updated the scaling relation to include pseudo bulges, using the database of \citet{KormendyHo2013}. For convenience, we repeat it here: $\log \left( \frac{M_{\rm BH}}{M_{\sun}} \right) = (-0.68 \pm 0.05) + (5.20 \pm 0.37) \log \left( \frac{\sigma}{200\; {\rm km\; s}^{-1}} \right)$. This relation has an intrinsic scatter of 0.44~dex.  We chose not to use the AGN luminosities derived through SED fitting to estimate $L_{\rm bol}$, because the typically low AGN fractions ($\lesssim 20\%$) of our sources to the SED render the AGN luminosities from SED fitting highly uncertain \citep{Ciesla2015}.

The distribution of Eddington ratio as a function of $M_*$ and sSFR (Figure~\ref{fig:AGN_Ms}) shows that AGNs in SFGs generally have more active BHs than those in quiescent hosts. At the same time, AGN hosts with higher sSFRs also accrete more vigorously, consistent with the positive correlation between the BH accretion rate and global SFR and SFE \citep{Zhuang2021, Zhuang2022}. In contrast to the star-forming AGNs, no statistically significant relation between Eddington ratio and stellar mass is seen in QGs. Gas-rich QGs have moderately higher Eddington ratios than gas-poor QGs, but the $\sim 0.4$~dex offset is within the $3\,\sigma$ uncertainty of the two populations. The majority of AGNs in QGs are low-luminosity systems with $L_{\rm bol}/L_{\rm Edd} \approx 10^{-5}$ to $10^{-3}$, a regime in which a radio-emitting jet or wind is commonplace \citep{Ho2002} and becomes increasingly energetically important \citep{Ho1999, Ho2008}. Jet-mode or kinetic-mode feedback may play a particularly critical role during the late phases of galaxy evolution (e.g., \citealt{McNamara2012, Weinberger2017, Yuan2018}). Radio jets can inhibit star formation by shock heating the gas \citep{Murthy2019, Krause2023}, or if interactions between the jet and the interstellar medium drive turbulence and outflows \citep{Mukherjee2018b, Murthy2019}. To assess the incidence of jets in our sample, we search for radio counterparts by cross-match our sample with the deep VLA 1.4~GHz observations of the Stripe~82 region performed by \cite{Hodge2011}\footnote{\url{http://www.physics.drexel.edu/~gtr/vla/stripe82/Deep\_VLA\_Observations\_of\_SDSS\_Stripe\_82.html}}. Matching the sky position within 5\arcsec\ ($\sim 3$ times the resolution of the radio synthesized beam) yields only a small handful of detections with $\rm S/N > 5$ (Table~\ref{tab_statistics}), far too few to yield any meaningful conclusions.

In the nearby Universe, AGNs exert a minimal instantaneous impact on the gas content and star formation activity of active galaxies (e.g., \citealt{Shangguan2018, Yesuf2020, Molina2022, Ward2022}). Instead, AGN feedback may inhibit gas cooling and thereby suppress star formation through cumulative effects \citep{Piotrowska2022, Bluck2023}. Within this framework, gas-rich QGs, relative to their gas-poor counterparts, can be regarded as systems that have undergone less exposure to the long-term effects of integrated AGN feedback. This expectation is not borne out in our sample, however. Using the mass of the central supermassive BH as a proxy for the total integrated energy from AGN feedback \citep{Goubert2024}, we perceive no notable difference between the BH masses of the two groups of QGs (Figure~\ref{fig:AGN_Ms}, bottom panels). As expected, at fixed stellar mass when $M_{*}>10^{10.5}\,M_\odot$ quiescent systems have higher $M_{\rm BH}$ than SFGs, suggestive of quenching at higher redshift from integrated AGN feedback.
The inverse correlation between $M_{\rm BH}$ and sSFR follows trivially from the variation of SFR with $M_{\rm *}$, and thus $M_{\rm BH}$ \citep{Magorrian1998, Greene2020}.

\begin{figure*}[t]
\centering
\includegraphics[width=0.95\textwidth]{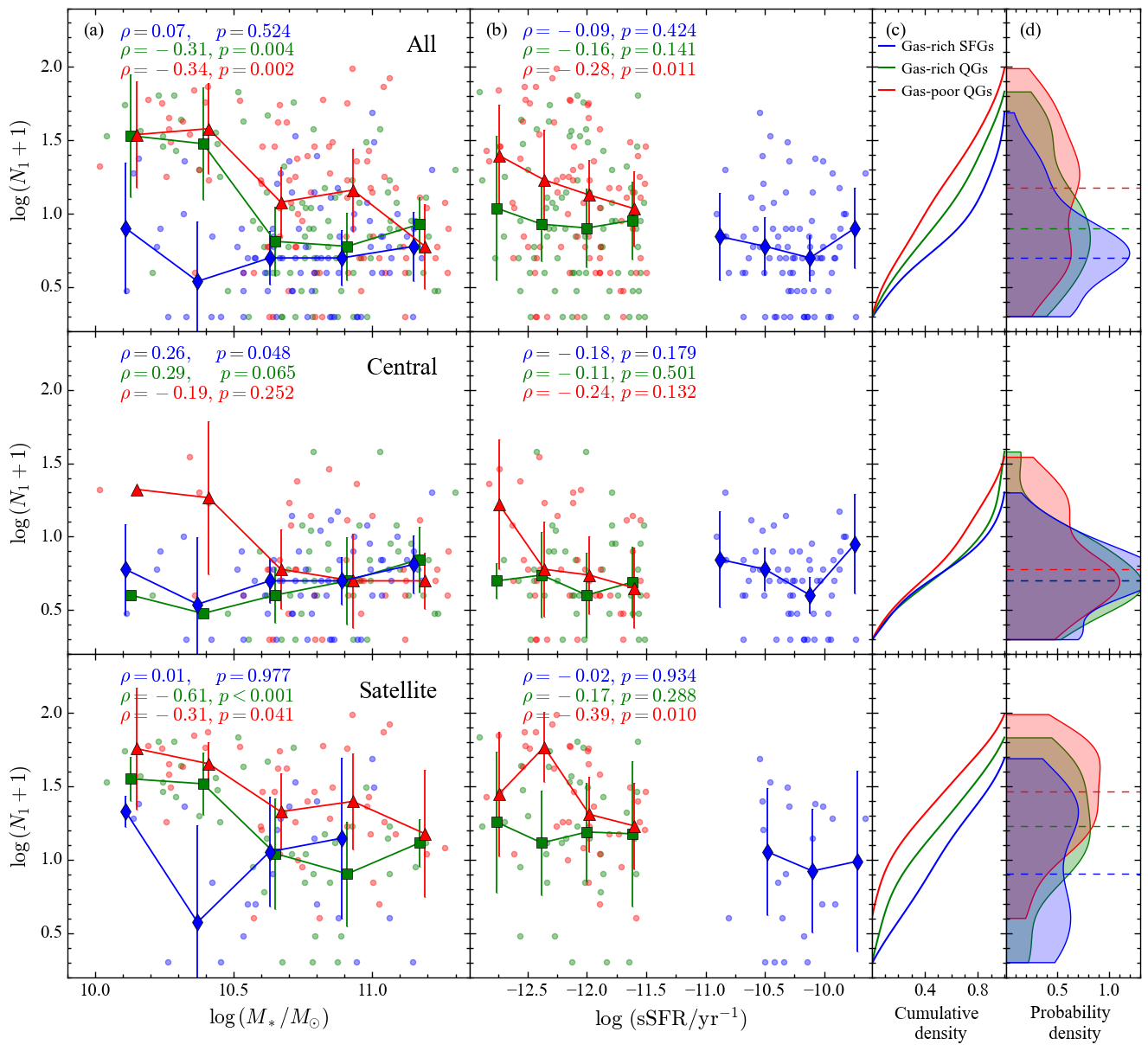}
\caption{The local environment density $N_1$ as a function of (a) $M_*$ and (b) sSFR for all (top) gas-rich SFGs (blue), gas-rich QGs (green), and gas-poor QGs (red), and then separately for those that are classified as central (middle) and satellite (bottom) galaxies. The normalized cumulative density and Gaussian-kernel probability density function for each sample are given in panels (c) and (d), respectively. Other conventions are similar to Figure~\ref{fig:n_total}.}
\label{fig:env_N1}
\end{figure*}

\begin{figure*}[t]
\centering    
\includegraphics[width=0.95\textwidth]{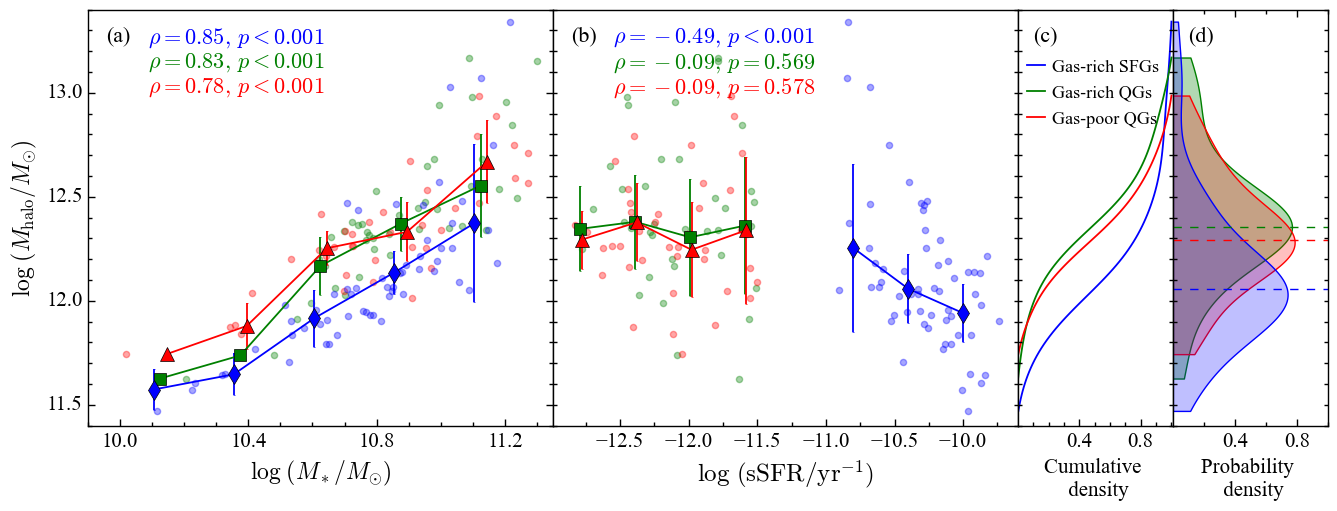}
\caption{The dark matter halo mass $M_{\rm halo}$ as a function of (a) $M_*$ and (b) sSFR for gas-rich SFGs (blue), gas-rich QGs (green), and gas-poor QGs (red). The normalized cumulative density and Gaussian-kernel probability density function for each sample are given in panels (c) and (d), respectively. Other conventions are similar to Figure~\ref{fig:n_total}.}
\label{fig:halo_Ms}
\end{figure*}

\vspace{1cm}
\subsection{Environmental Effects}\label{sec_environment}

Regardless of the still-unclear internal mechanisms of suppressing star formation (Sections~\ref{sec_morphology}--\ref{sec_AGN}), we have yet to answer an even more basic question. Where did the cold gas in QGs come from?  Was it acquired externally through mergers with neighboring galaxies?  Or was it accreted from the halo? 

Figure~\ref{fig:env_N1} explores these issues by examining the local density $N_1$, which represents the number of neighboring galaxies within a radius of $1\,$Mpc and a radial velocity difference $\Delta v < 1000\,\rm{km\,s^{-1}}$ around each galaxy of interest \citep{He2022}. We calculate $N_1$ using the galaxies in the SDSS database.  Only galaxies with absolute $r$-band magnitude brighter than $-19.5$ are included in the analysis. The target galaxy itself is also included in the number count if it is brighter than the magnitude cut-off. The $N_1$ parameter can be translated approximately into an indicator of the environment on larger scales. Similar to \cite{He2022}, field galaxies correspond to $N_1<3$, galaxy groups $4\leq N_1 \leq 10$, and large galaxy groups and clusters $N_1 > 10$. We adopt these conventions, and in the following discussion we consider a galaxy to reside in a cluster if $\log\, (N_1+1)  \gg 1$. 

Looking first at all galaxies irrespective of their environments (top row of Figure~\ref{fig:env_N1}), the median local density of the gas-poor QGs ($N_1 = 14^{+28}_{-11}$) appears intriguingly higher than that of the gas-rich QGs ($N_1 = 7^{+26}_{-5}$), particularly at the high-$M_*$ and low-sSFR end.  Both have higher local density than the SFGs ($N_1 = 4^{+7}_{-3}$). The QG samples show a weak inverse correlation between local density and $M_*$ and sSFR, possibly because massive galaxies are more likely to have undergone multiple merger events, which have removed their low-mass neighbors. 

The relative differences in local density among the three comparison populations become more evident after sorting the samples into central or satellite galaxies according to the SDSS group catalog of \cite{YangXH2007}, as illustrated in the middle and bottom rows of Figure~\ref{fig:env_N1}. The star-forming sample occupies $71\%\pm 12\%$ of the central galaxies, higher than the percentage of central galaxies in the gas-rich ($51\%\pm 10\%$) and gas-poor ($48\%\pm 9\%$) quiescent samples. The central galaxies exhibit a comparable local density across the three samples: $N_1 = 4^{+4}_{-3}$ for the SFGs, $N_1 = 4^{+5}_{-2}$ for the gas-rich QGs, and $N_1 = 5^{+11}_{-3}$ for the gas-poor QGs. By contrast, satellite galaxies behave markedly differently, showing median $N_1 = 7^{+16}_{-5}$ for SFGs, $N_1 = 16^{+26}_{-10}$ for gas-rich QGs, and $N_1 = 28 ^{+30}_{-18}$ for gas-poor QGs. At the high-$M_*$ and low-sSFR end, gas-poor QGs reside in substantially denser environments than gas-rich galaxies, with a difference that exceeds the $3\,\sigma$ uncertainties of the median values.

\subsection{Dark Matter Halo Quenching}\label{sec_halo}

To investigate the possible dependence of gas content on the mass of the dark matter halo, we use the catalog of halo masses of \cite{ZhaoDY2024}, who developed a new machine-learning method to predict the halo mass from multiple physical properties of galaxy groups. As illustrated in Figure~\ref{fig:halo_Ms}, both types of QGs possess similar halo masses. The halo mass has little bearing on the origin or disposition of the interstellar medium in QGs. At the same $M_*$, SFGs inhabit halos of distinctly lower mass than the QGs. We note that the majority of the galaxies in all three samples---including the SFGs---reside in halos exceeding the mass threshold of $M_{\rm halo} = 10^{12}\,M_{\odot}$ that is thought to be critical for the shock-heating of cold gas in the circumgalactic medium \citep{Dekel2006}. This reflects the manner in which we constructed the comparison samples for the gas-rich QGs (see Section~\ref*{sec_sample} and Figure~\ref{fig:hist0}). As a consequence of the tight correlation between the stellar mass and halo mass \citep{Conroy2009, Moster2013, Wechsler2018}, sSFR correlates inversely with halo mass in SFGs, but not in QGs. In the QG samples, the uncertainties of the SFRs can reach $\sim 0.6$ dex, flatting the trend between $M_{\rm halo}$ and sSFR. Evidently, an elevated halo mass is not a sufficient condition for halting star formation.

\section{Implications}\label{sec_discussion}

A principal and somewhat unexpected finding of this work is that the internal physical properties of quiescent galaxies have little bearing on their gas content. This holds true for their global morphology, as measured by their \sersic\ index, light concentration, or bulge-disk structure. The majority ($\gtrsim 70\%$) of QGs are bulge-dominated disk systems, consistent with the notion that a large fraction of early-type galaxies have significant rotation \citep{Bundy2010, Emsellem2011, Bell2012}. Gas-rich and gas-poor QGs follow the same stellar mass-size relation, indicating that both groups have undergone similar morphological transformations. They also have similar dark matter halo masses.

We find no evidence either that the mass of the BH or the presence of the AGN has any bearing on the gas content in QGs. Most of the AGNs in our QGs have low luminosity, suggesting that jet-mode feedback plays an increasingly important role in them \citep{Ho2002, Ho2008}.  In low-luminosity AGNs, jet-mode feedback can interact with the gas of the host galaxy in some specific situations. For instance, the jet can sweep up gas from the central regions in cases where the radio jet is not perpendicular to but inclined toward the galactic plane \citep{Combes2017, Cielo2018, Mukherjee2018a}. However, gas-rich QGs have only marginally ($<3\sigma$) higher Eddington ratios than their gas-poor counterparts, in all the stellar mass or sSFR bins (Figure~\ref*{fig:AGN_Ms}). This is expected because gas content correlates not only with SFR but also the BH accretion rate \citep{Zhuang2020, Zhuang2021}. Indeed, it may seem counterintuitive that the SFG sample has a higher fraction of low-luminosity AGN (mainly type 2 AGNs and composites) hosts than the QG sample, and that the low-luminosity AGNs hosted by SFGs, if anything, generally have even higher Eddington ratios than those that reside in QGs (see Figure 1 in \citealt{Leslie2016}). This is consistent with the observation that the BHs in early-type galaxies, which dominate the QG samples in this work, have weaker AGN activity than those in late-type galaxies \citep{Ho2009a}.  Our results disagree with galaxy quenching by instantaneous AGN feedback. In contrast, the positive correlation between the Eddington ratio of AGNs and sSFR in stellar mass-matched samples (Figure~\ref{fig:AGN_Ms}b) corroborates the argument that positive AGN feedback may boost star formation \citep{Bernhard2016, Zhuang2020, Salome2015}.  

Our results imply that AGN feedback is ineffective in blowing the gas out of galaxies or suppressing their SFE, irrespective of whether they are passive or forming stars. This qualitatively agrees with the evidence that the kinetic power associated with the cold gas outflows in radio galaxies is much less than required to halt star formation \citep{Morganti2010}. Even in the galaxies with intense BH activity, such as quasars, the AGN feedback that is expected to be dominated by the radiative mode seems to be ineffective in expelling the gas out on galactic scales \citep{Shangguan2018, Shangguan2019, Molina2022}. 

Among the physical properties examined in this paper, the most notable distinction between gas-rich and gas-poor QGs lies in the local environmental density. As shown in Figure~\ref*{fig:env_N1}, the environment affects the gas content of the central and satellite QGs differently. Moreover, the gas content of QGs differs for stellar masses above and below $M_{*}\approx 10^{10.5}\,M_{\odot}$. 

As shown in the first row of Figure~\ref*{fig:env_N1}a, the gas-rich QGs in the high-mass end of the distribution reside in environments with a density comparable to that of the SFGs but lower than that in which gas-poor galaxies reside. This difference mainly comes from the contribution of satellite galaxies (bottom row of Figure~\ref*{fig:env_N1}a). In contrast, the central QGs (both gas-rich and gas-poor) share the characteristics of the SFGs in the massive end, with $\sim 4$ bright neighbors on average within a radius of $1\,\rm Mpc$, resembling small galaxy groups \citep{He2022}. Interestingly, \citet{LiXiao2024} find that passive \HI-rich galaxies, characterized by high \HI-to-stellar mass ratios, in contrast to passive \HI-normal galaxies, tend to be associated with lower density environments with fewer neighbors on scales up to a few Mpc. Compared to the randomly selected bulge-dominated early-type galaxies, \citet{LiFujia2024} find that the \HI-rich bulge-dominated early-type galaxies are located in lower density surroundings and show lower quiescent fractions of their neighbors within $2\,\rm Mpc$. The lower-density environment and more star-forming neighbors provide a potential \HI-rich reservoir, which is a necessary condition for gas accretion from a nearby environment. However, most of their bulge-dominated early-type galaxies are central galaxies.
We find similar but not identical results: only satellite QGs show a significant difference in their environments. \citet{LiXiao2024} further suggested that the passive \HI-rich galaxies may be located preferentially at the center of relatively low-mass halos, but we see no connection between gas content and halo mass among central galaxies (Section~\ref{sec_halo}). 

Why might gas-rich satellite QGs reside preferentially in lower-density environments than their gas-poor counterparts? Massive satellite galaxies in clusters at $z = 0$ may have been pre-processed and quenched before they fell into clusters at higher redshifts \citep{Mihos2004, Vijayaraghavan2013, Gabor2015}. Upon entering a cluster, massive QGs in a relatively sparser environment can protect their gas---if any remains---from ram pressure stripping or tidal truncation. This arises from the fact that with increasing stellar mass, the temperature and pressure differences between the satellite’s subhalo and its host halo are smaller and the potential well of the satellite is deeper, making the stripping of the satellite’s subhalo inefficient and gas accretion easier \citep{VanDeVoort2017}. Recent gas accretion is another possible origin of the gas-rich QGs living in a relatively low-density environment. The higher incidence of counter-rotation and kinematic misalignment found in \HI-rich galaxies with low levels of star formation \citep{Sharma2023} suggests that such systems may have been created by gas infall onto gas-poor galaxies \citep{Starkenburg2019}. The efficiency of gas accretion decreases significantly toward the cluster center, where the environmental density is high \citep{VanDeVoort2017}. The central regions are too remote from the intragalactic material to trigger cold flow accretion \citep{Sharma2023}. In addition, in low-density environments, QGs have a higher probability of gas replenishment by gas-rich mergers on account of the lower velocity dispersion among galaxies in these regions \citep{Mihos2004, Fakhouri2010, Belli2021}. Gas-rich minor mergers can lower the SFE of the primary galaxies \citep{Li2023ApJ}.
 
Having acquired or retained their gas, how do gas-rich satellite QGs remain quiescent?  We surmise that their prominent bulges, as indicated by the large $n$ and $C$, elevate the Toomre $Q$ parameter and hence suppress disk fragmentation and thus star formation. On the other hand, in galaxies with massive bulges, strong integrated AGN feedback, as indicated by $M_{\rm BH}$  (see bottom panels of Figure~\ref{fig:AGN_Ms}), helps keep the host galaxies quiescent by ionizing and heating the interstellar medium in the vicinity of the nucleus \citep{Jin2024}. A low SFE may also be a consequence of the low efficiency of converting \HI\ to $\rm H_2$ owing to the low column density or turbulence induced by accretion in the disk outskirts \citep{Bigiel2008, Forbes2023}. The stronger shear can reshape the gas distribution \citep{Lu2024} to further increase Toomre $Q$.

In the presence of non-axisymmetric gravitational perturbations from a bar or spiral arms, the newly accreted gas in the disk can be funneled to the center and consumed by star formation. The initially gas-rich QGs will evolve into gas-poor ones. However, the bars can effectively drive gas inflow only if it reside within the corotation radius of the bar \citep{Kuno2007, Athanassoula2013}. If the angular momentum of the accreted gas is high, it will reside in the outer disk for a long time in the absence of further external perturbations to remove its angular momentum \citep{Peng2020, Renzini2020}. The long timescale of secular evolution in such galaxies may partially explain the existence of gas-rich QGs (Section~\ref*{sec_bar}).  

In the low-mass end ($M_* = 10^{10}-10^{10.5}\,M_{\odot}$), the majority of satellites in high-density environments are quenched, consistent with the environmental quenching scenario \citep{Davies2016, Davies2019}. Overdense environments tend to remove or inhibit the supply of gas in satellite galaxies of low to intermediate mass, by mechanisms such as starvation/strangulation \citep{Moore1999, Peng2015}, tidal and ram pressure stripping \citep{Gunn1972, Moore1999, Brown2017}, or harassment \citep{Moore1996}. To remain quiescent, low-mass satellites should be gas-poor. However, we find that low-mass satellites can still retain a significant amount of cold gas even in a dense environment. The low-mass quiescent satellites exhibit similar environments for both the gas-rich and gas-poor samples, indicating that environmental density is not the only major factor influencing the gas content in these satellite galaxies. Ultimately, we are unable to elucidate the reason for the high gas content observed in these galaxies. Perhaps the interstellar medium in low-mass systems is replenished largely by recycled gas from internal stellar mass loss instead of external gas accretion \citep{Segers2016}. However, outflows driven by stellar feedback are more effective in the shallower gravitational potential of low-mass galaxies. \citet{Trussler2020} proposed that different outflow strengths and the recycling timescales of the escaped gas may result in QGs with varying gas content.

\section{Conclusions}\label{sec_summary}

In this study, we compare the fundamental physical properties of gas-rich and gas-poor quiescent and star-forming galaxies, ensuring that their stellar mass, SFR, and gas mass are matched. This approach allows us to investigate the formation mechanisms of gas-rich QGs in a controlled manner. 
Our main conclusions are summarized as follows. 

\begin{itemize}
    \item The gas-rich and gas-poor QGs occupy similar distributions in the mass-size relation. The two QG samples are in a similar evolutionary stage, which is considerably more evolved than the SFGs.

    \item QG samples have a higher fraction of early-type disk morphology and higher morphological concentration than the SFGs, but no morphological difference is found between the gas-rich and gas-poor QG samples.

    \item The gas-rich and gas-poor QGs have comparable AGN strength, BH mass, and dark matter halo mass, which are not responsible for the difference in gas content in the QGs.

    \item The main difference between the gas-rich and gas-poor QGs is the local environmental density, primarily for the massive satellites. Gas-poor satellite QGs are mostly located in large groups or clusters, while their gas-rich counterparts are in small groups or away from the dense cluster centers. Central QGs with different gas content do not show distinct environmental densities. We suggest that the gas-rich satellite QGs originate from the accretion or conservation of cold gas in the low-density environment.

    \item The significant bulge components that stabilize the gas disks could be responsible for the low SFE in the gas-rich QGs. Integrated AGN feedback, if present, could help maintain the bulge-dominated host galaxies quiescent. The QGs with accreted gas residing in the outer disks exhibit a long timescale of gas infalling, keeping them at low SFE.  
    
\end{itemize}

\newpage
\begin{acknowledgments}
This work is supported by the National Key R\&D Program of China (2022YFF0503401, 2024YFA1611602), the National Natural Science Foundation of China (No. 11991052, 12233001, 12011540375), the Shanghai Natural Science Research Grant (24ZR1491200), the ``111'' project of the Ministry of Education under grant No. B20019, and the China Manned Space Project (CMS-CSST-2021-A04, CMS-CSST-2021-A06). YAL is funded by the China Postdoctoral Science Foundation (No. 2023M742285) and the Postdoctoral Fellowship Program of CPSF under Grant Number GZC20231611.  Y.P. acknowledges support from the National Science Foundation of China (NSFC) grant Nos. 12125301, 12192220, 12192222, and support from the New Cornerstone Science Foundation
through the XPLORER PRIZE. We thank D. Donevski, G. Lorenzon, Yuanqi Liu, and Jing Wang for discussions and comments on this work. We are grateful to Dingyi Zhao for sharing the catalog of halo masses of \cite{ZhaoDY2024} in advance of publication.

\end{acknowledgments}

\end{document}